\documentclass[
 aps, pra,
 amsmath,amssymb,
 11pt,
 final,
tightenlines,
 twoside,
 twocolumn,
 nofloats,
nofootinbib,
 superscriptaddress,
showkeys,
showkeywords,
 ]
{revtex4-2}

\usepackage[T2A]{fontenc}
\usepackage[utf8x]{inputenc}
\usepackage[english]{babel}
\usepackage{graphicx}
\usepackage{dcolumn}
\usepackage{bm}

\input{maik.rty}

\setcitestyle{authoryear,round,aysep={,},citesep={;}}
\def\squareforqed{\hbox{\rlap{$\sqcap$}$\sqcup$}}

\def\sq{\ifmmode\squareforqed\else{\unskip\nobreak\hfil
\penalty50\hskip1em\null\nobreak\hfil\squareforqed
\parfillskip=0pt\finalhyphendemerits=0\endgraf}\fi}

\def\utw{\smash{\rlap{\lower5pt\hbox{$\sim$}}}}

\def\udtw{\smash{\rlap{\lower6pt\hbox{$\approx$}}}}

\def\diameter{{\ifmmode\mathchoice
{\ooalign{\hfil\hbox{$\displaystyle/$}\hfil\crcr
{\hbox{$\displaystyle\mathchar"20D$}}}}
{\ooalign{\hfil\hbox{$\textstyle/$}\hfil\crcr
{\hbox{$\textstyle\mathchar"20D$}}}}
{\ooalign{\hfil\hbox{$\scriptstyle/$}\hfil\crcr
{\hbox{$\scriptstyle\mathchar"20D$}}}}
{\ooalign{\hfil\hbox{$\scriptscriptstyle/$}\hfil\crcr
{\hbox{$\scriptscriptstyle\mathchar"20D$}}}}
\else{\ooalign{\hfil/\hfil\crcr\mathhexbox20D}}%
\fi}}

\usepackage{adjustbox}

\begin{document}



\title{Study of the $\mu$\,Cet Binary with Speckle Interferometric, Photometric and Spectroscopic Techniques}

\author{\firstname{V.~V.}~\surname{Dyachenko}}
 \email{dyachenko@sao.ru}
 \affiliation{Special Astrophysical Observatory,  Russian Academy of Sciences, Nizhnii Arkhyz, 369167 Russia}
\author{\firstname{I.~A.}~\surname{Yakunin}}
\email{elias@sao.ru}
 \affiliation{Special Astrophysical Observatory,  Russian Academy of Sciences, Nizhnii Arkhyz, 369167 Russia}
\author{\firstname{R.~M.}~\surname{Bayazitov}}
 \affiliation{Kazan (Volga Region) Federal University, Kazan, 420008 Russia}
\author{\firstname{S.~A.}~\surname{Grigoriev}}
 \affiliation{Kazan (Volga Region) Federal University, Kazan, 420008 Russia} 
\author{\firstname{T.~A}~\surname{Ryabchikova}}
 \affiliation{Institute of Astronomy, Russian Academy of Sciences, Moscow, 119017 Russia}
\author{\firstname{Yu.~V.}~\surname{Pakhomov}}
 \affiliation{Institute of Astronomy, Russian Academy of Sciences, Moscow, 119017 Russia}
\author{\firstname{E.~A.}~\surname{Semenko}}
\affiliation{National Astronomical Research Institute of Thailand, Mae Rim, Chiang Mai 50180, Thailand}
\author{\firstname{A.~S.}~\surname{Beskakotov}}
 \affiliation{Special Astrophysical Observatory,  Russian Academy of Sciences, Nizhnii Arkhyz, 369167 Russia}
\author{\firstname{A.~A.}~\surname{Mitrofanova}} 
\affiliation{Special Astrophysical Observatory,  Russian Academy of Sciences, Nizhnii Arkhyz, 369167 Russia}
 \author{\firstname{A.~F.}~\surname{Maksimov}}
 \affiliation{Special Astrophysical Observatory,  Russian Academy of Sciences, Nizhnii Arkhyz, 369167 Russia} 
 \author{\firstname{Yu.~Yu.}~\surname{Balega}}
 \affiliation{Special Astrophysical Observatory,  Russian Academy of Sciences, Nizhnii Arkhyz, 369167 Russia}
 
\begin{abstract}
We present a refined speckle-interferometric orbit of a binary system $\mu$\,Cet, with the main component studied based on the analysis of photometric and spectroscopic data, obtained at the SAO RAS 6-m telescope. The object was initially classified as a giant with chemical composition anomalies.  As a result of our analysis, we conclude that the star belongs to the Main Sequence, to the class of non-peculiar stars. Analysis of photometric data from the TESS mission indicates that the main component of the system belongs to the $\gamma$\,Dor pulsators.
\end{abstract}

\maketitle

\section{INTRODUCTION}

Binary stars are a natural means of studying the physical evolution of stars and measuring their mass, the most important fundamental parameter. The systems that include objects with distinct features, namely, variables, peculiar and evolving stars are unique laboratories for reliably determining the physical state, preconditioning the observed manifestations of these objects. 
Among the tight pairs, we have identified a number of objects suitable for a speckle interferometric study  at the 6-m telescope of the Special Astrophysical Observatory of the Russian Academy of Sciences (SAO RAS). 
One of them is the $\mu$\,Cet system (HD\,17094), a speckle interferometric binary star, the main component of which was suspected by Hauck (1971) to belong to the $\delta$\,Scuti variables. At that, a later classification (Gray et al., 2003) grades the bright component as a giant of spectral class A9 with strengthened metal lines.

The $\mu$\,Cet system is known as a type SB1 spectroscopic binary. The spectroscopic orbital solution was published by Abt~(1965). Its period is 1202.2~days or 3.29~years. It was noted that the obtained curve may be incorrect. Later, Abt and
Levy (1974) reported a probable constant radial velocity. However, this star is included in The Ninth Catalogue of Spectroscopic Binary Orbits (SB9) (Pourbaix et al., 2004) with a low-confidence solution.

It was first resolved interferometrically in 1982~and got the designation TOK1 (Tokovinin, 1985). We have previously constructed a preliminary orbit of the speckle interferometric pair $\mu$\,Cet (Dyachenko et al., 2019).

According to the Gaia Collaboration (2020), the distance to the star is 26~pc, based on the ESA data (1997), the stellar magnitude is \mbox{$m_V=4\,.\!\!^{\rm m}3$}.  The proximity of the star and its brightness create all the conditions for the accurate photometric and spectroscopic analysis, since the interstellar absorption can be neglected (Lallement et al., 2022) and spectra with a high
$S/N$ ratio can be obtained on the spectrographs with a high resolving power $R=\lambda/\Delta\lambda$
(see Section~2.2). However, only one high-resolution spectrum was found in the archives of spectroscopic observations (the Elodie spectrum, $R=42\,000$), but it has a low $S/N=88$.

In the present study, we have constructed the orbit of the binary $\mu$\,Cet. The contribution of the secondary (B) component to the photometric observations was estimated. The atmospheric parameters of the primary (A) component of the $\mu$\,Cet system:  $T_{\rm eff}$, $\log g$, microturbulent velocity $\xi_{t}$, metallicity $[M/{\rm H}]$ and the projection of the rotation velocity onto the line of sight $v\sin i$ were refined using a joint analysis of photometric and spectroscopic observations. 
An upper limit on the magnetic field was obtained using the polarization observations. 
In the process of a detailed spectroscopic analysis, we determined the abundances of 14~chemical elements, including  such elements as Si, Cr, Sr, Eu, which reveal the abundance anomalies in peculiar stars. The estimates of the age of $\mu$\,Cet and the masses of the system components were obtained.

\section{OBSERVATIONS AND DATA REDUCTION}
\subsection{Speckle Interferometric Observations}\label{Obssi}

The observations using the speckle interferometry technique  (Labeyrie, 1970) were carried out at the SAO RAS  6-m telescope over 2011--2023. We used a speckle interferometer (Maksimov et al., 2009) based on the  Andor iXon Ultra 897 EMCCD. 
The images were registered in series of 2000 frames with 20-ms exposures.  
The interference filters (central wavelength/half-width) used were: 550/20, 550/50, 694/10 and 800/100 nm.
This paper contains the previously unpublished 19 series, covering 16 observation epochs of \mbox{2018--2023}.

Preliminary processing and analysis of the series were performed by parameter selection in the ring regions of the power spectrum (Balega et al., 2002; Pluzhnik, 2005; Obolentseva et al., 2021). For each series, a set of positional parameters $\rho$, $\theta$ was obtained and the brightness difference between the  $\Delta m$ components was determined.

The images were reconstructed using the bispectral analysis method (Lohmann et al., 1983). An example of a reconstructed image is shown in Fig.~1.

Preliminary estimates of the orbital parameters were calculated using the Monet method (Monet, 1977). The resulting parameters were determined using a software package based on the {\tt ORBIT} (Tokovinin, 1992).

\begin{figure}	
\includegraphics[scale=0.4]{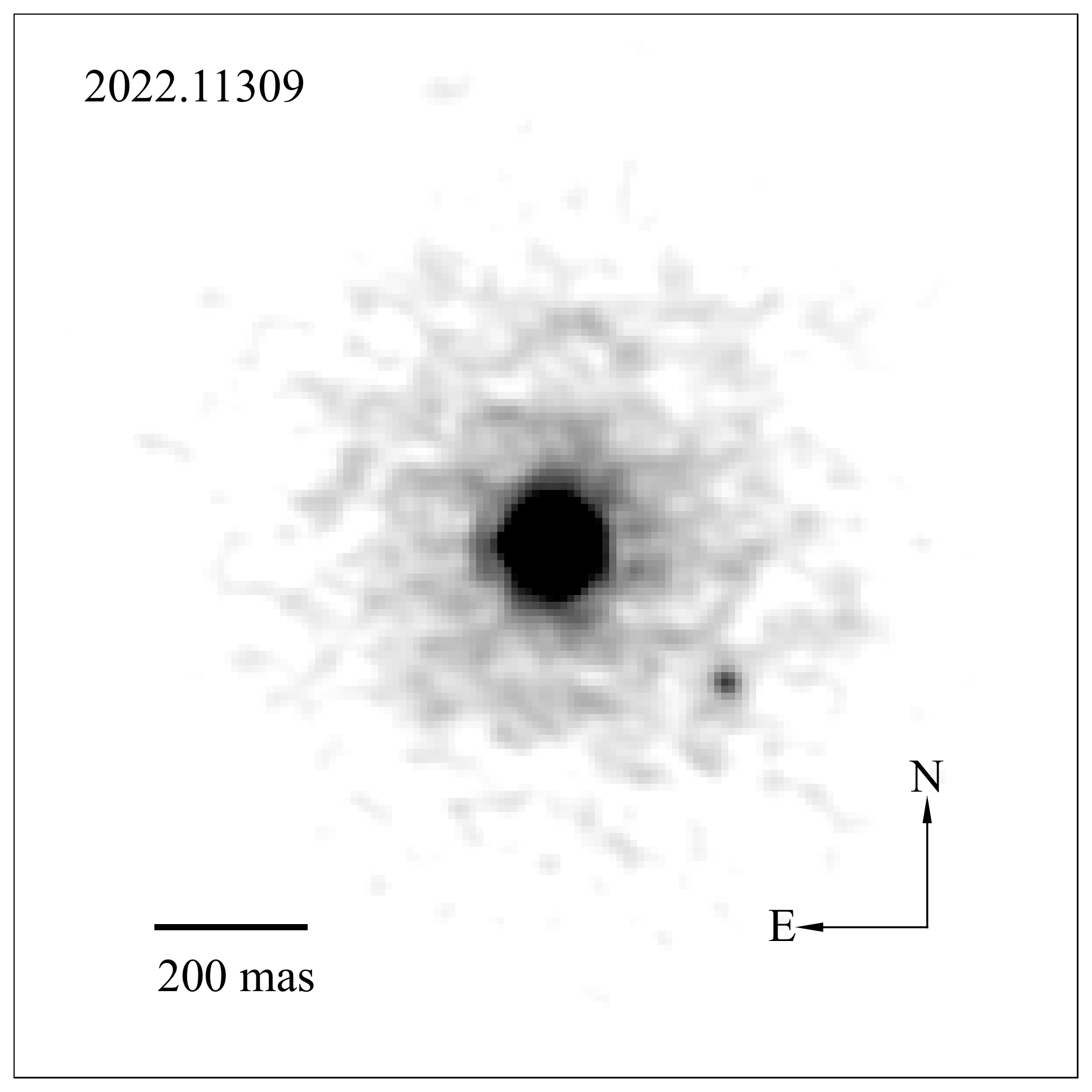}
	\caption{The image of the $\mu$\,Cet system, reconstructed applying the bispectral analysis. Epoch 2022.11309, a 550/50~nm filter.}\label{fig:Rec2023}
\end{figure}

\subsection{Spectroscopic Observations}\label{Obssp}

For the spectroscopic analysis, we used the spectrum of $\mu$\,Cet from the open  ELODIE Echelle Spectrometer Archive with \mbox{($S/N=88$}, $R=42\,000$, the observed range of \mbox{4000--6800\,\AA}), obtained on September 25, 1994~ on the 193-cm reflector at the Haute-Provence Observatory\footnote{\url{http://atlas.obs-hp.fr/elodie/}}. We normalized the spectrum to the continuum level ourselves. In doing so, special attention was paid to the regions containing the H${\alpha}$ and H${\beta}$ hydrogen lines, since they play a key role in determining the parameters of the stellar atmosphere.

To estimate the magnetic field, we used the circularly polarized spectra (\mbox{$R=15\,000$}, $S/N=150$) obtained with the Main Stellar Spectrograph (MSS)\footnote{MSS, \tt https://www.sao.ru/hq/lizm/mss/en/} of the 6-m SAO RAS BTA telescope on March 8, 2019 in blue (4430--4987\,\AA) and red ($5860$--$6700$\,\AA) filters. 
The primary reduction of the spectrum was carried out using the {\tt ESO MIDAS} software package (see Kudryavtsev, 2000).
The unpolarized spectrum of the star was obtained by summing the circularly polarized spectra.

\subsection{TESS Mission Photometric Data}\label{Obstess}

\begin{figure}
	\includegraphics[width=\linewidth]{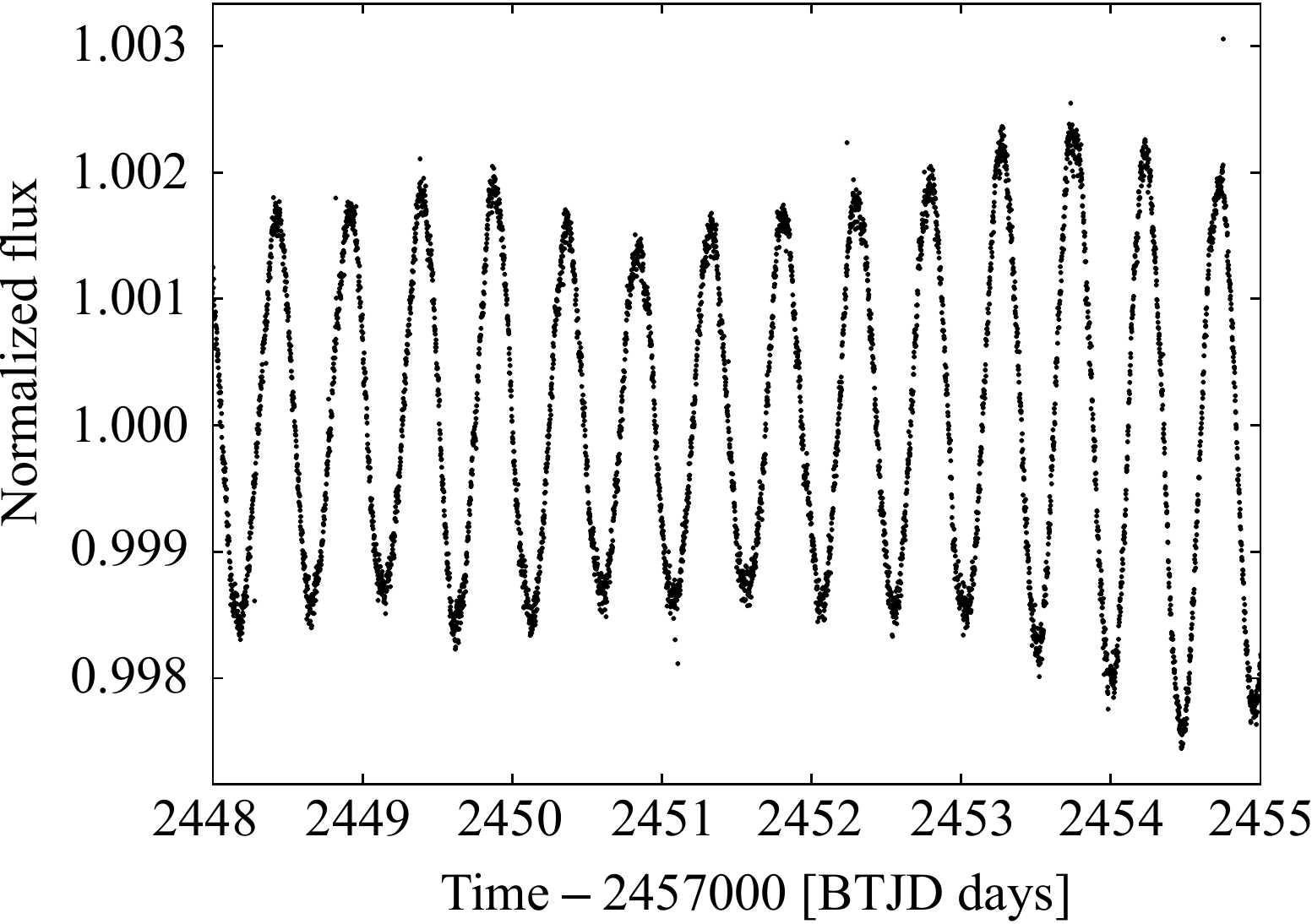}
	\caption{A part of the light curve of $\mu$\,Cet according to the TESS mission data.}\label{tesslc}
\end{figure}

During the TESS space mission\footnote{The Transiting Exoplanet Survey Satellite: {\tt https://exoplanets.nasa.gov/tess/}} (Ricker
et al., 2015), photometric observations with two-minute exposures were obtained for more than 200\,000 objects from the Northern and Southern Hemispheres (Stassun et al., 2018). 
The telescope is equipped with four cameras with a field of view of $24^\circ\times24^\circ$, giving an effective field of view of $24^\circ\times96^\circ$ with a scale of $21^{\prime\prime}$/pixelы. Each such section, a sector, is observed for 27.4~days,  every 13.7~days the observations are suspended for 16~hours to transmit the data to ground stations. 
Then the process is reiterated for the following sector. 
There are 13~sectors in each hemisphere and thus, the telescope covers 85\% of the entire sky over two years of observations.
The telescope is equipped with a 600--1000-nm passband filter, 
with a central wavelength of 786.5 nm, corresponding to the $I$ band in the Johnson-Cousins system.
Photometric data for individual sectors are presented in the form of Target Pixel Files (TPF), the images of the object, from which, using the SPOC (the TESS Science Processing Operations Center) technique \footnote{\url{https://archive.stsci.edu/hlsp/tess-spoc}} (Jenkins et al., 2016), three types of observational time series are extracted: SAP (Simple Amplitude Aperture) and PDCSAP (Pre-search Data Conditioning SAP) fluxes, as well as a background signal. 
The SAP flux values are extracted from an aperture, selected to optimize the $S/N$ ratio, and are then corrected for the measured background. 
This approach may leave some systematic effects, such as those introduced by the satellite spin. Such systematics is minimized in the PDCSAP data by means of the Co-trending Basis Vectors (CBVs). In this paper we use the PDCSAP flux values.

The $\mu$\,Cet system was observed by the TESS telescope in sectors 42, 43, 44, and 70~with  two-minute exposures. A fragment of the light curve of sector~42 is shown in Fig.~2. We analyzed each sector separately, as well as the combined data from sectors 42, 43, 44 and sectors 42, 43, 44, and 70. 
Frequency, amplitude and phase measurements were made using the  {\tt Period04} code (Deeming, 1975; Kurtz,
1985), which is based on the discrete Fourier transform. 
The extraction process continued until the condition $S/N > 4$ was satisfied (Breger et al., 1993). 
Note that there are also more complex stop criteria (see, for example, Van Reeth et al., 2015). 
The $S/N$ value at a given frequency was defined as the ratio of the peak amplitude on the power spectrum to the average value of  counts in a window of 1~day$^{-1}$. 
Two frequencies were considered unresolved under the condition $|\nu_i-\nu_j| < 2/T$, where $T$~ is the timed window of observations. Parameter errors were calculated using the method, described by Montgomery and O'Donoghue (1999).

The results of frequency measurements are given in Tables~1--3. The frequency $2.06$~day$^{-1}$ \mbox{($P = 0\,.\!\!^{\rm d}485$)} with the maximum $S/N$ ratio is highlighted in bold in all tables. Some of the frequencies obtained from the 42nd, 43rd, 44th and 70th sectors, on the one hand, may have no physical meaning, but be a consequence of the complex structure of the spectral window.
On the other hand, when comparing Table~3 with Tables~1--2, it is evident that all frequencies measured in sectors 42, 43, 44 (both separately and altogether) were also obtained by analyzing the combined data of sectors 42, 43, 44, and 70. The only exceptions are the frequencies of 2.1749~days$^{-1}$, according to the sector 44 data. It is fair to assume that in addition to 2.06~days$^{-1}$, $\mu$\,Cet exhibits variability with other periodicities.

\renewcommand{\baselinestretch}{0.75}
\setlength{\tabcolsep}{1.4pt}
\begin{table}[] 
\caption{Frequencies extracted for $\mu$\,Cet from the TESS photometric data. Here and below, the measurement error corresponding to the last significant digit of each parameter is indicated in the brackets} 
\medskip
\begin{tabular}{c|c|c|c}
\hline
\multicolumn{1}{c|}{Frequency,} & \multicolumn{1}{c|}{Amplitude,}   & \multicolumn{1}{c|}{\multirow{2}{*}{Phase}}               & \multirow{2}{*}{$S/N$}           \\ 
\multicolumn{1}{c|}{day$^{-1}$} & \multicolumn{1}{c|}{mmag}   & & \\
\hline
\multicolumn{4}{c}{Sector~42}  \\ 
\hline\multicolumn{1}{c|}{1.6497(9)}           & \multicolumn{1}{c|}{0.053(2)}          & \multicolumn{1}{c|}{0.56(1)}            & 4.9           \\
\multicolumn{1}{c|}{1.9322(3)}           & \multicolumn{1}{c|}{0.130(2)}          & \multicolumn{1}{c|}{0.156(2)}           & 5.0           \\
\multicolumn{1}{c|}{\textbf{2.06213(3)}} & \multicolumn{1}{c|}{\textbf{1.712(2)}} & \multicolumn{1}{c|}{\textbf{0.3009(1)}} & \textbf{42.8} \\
\multicolumn{1}{c|}{2.3131(4)}           & \multicolumn{1}{c|}{0.109(2)}          & \multicolumn{1}{c|}{0.457(3)}           & 6.3           \\
\multicolumn{1}{c|}{4.121(1)}            & \multicolumn{1}{c|}{0.044(2)}          & \multicolumn{1}{c|}{0.95(1)}            & 7.9           \\
\multicolumn{1}{c|}{4.245(2)}            & \multicolumn{1}{c|}{0.051(2)}          & \multicolumn{1}{c|}{0.69(1)}            & 5.9           \\ 
\hline
\multicolumn{4}{c}{ Sector~43}   \\ 
\hline
\multicolumn{1}{c|}{0.09562(4)}          & \multicolumn{1}{c|}{0.092(2)}          & \multicolumn{1}{c|}{0.758(3)}           & 4.7           \\
\multicolumn{1}{c|}{1.9412(3)}           & \multicolumn{1}{c|}{0.128(2)}          & \multicolumn{1}{c|}{0.958(2)}           & 5.6           \\
\multicolumn{1}{c|}{\textbf{2.05990(3)}} & \multicolumn{1}{c|}{\textbf{1.696(2)}} & \multicolumn{1}{c|}{\textbf{0.9209(2)}} & \textbf{32.9} \\
\multicolumn{1}{c|}{2.3064(4)}           & \multicolumn{1}{c|}{0.098(2)}          & \multicolumn{1}{c|}{0.952(3)}           & 7.3           \\
\multicolumn{1}{c|}{4.246(1)}            & \multicolumn{1}{c|}{0.036(2)}          & \multicolumn{1}{c|}{0.982(7)}           & 4.2           \\ 
\hline
\multicolumn{4}{c}{Sector~44} \\ \hline
\multicolumn{1}{c|}{\textbf{2.05989(3)}} & \multicolumn{1}{c|}{\textbf{1.344(2)}} & \multicolumn{1}{c|}{\textbf{0.8262(2)}} & \textbf{30.1} \\
\multicolumn{1}{c|}{2.1749(2)}           & \multicolumn{1}{c|}{0.195(2)}          & \multicolumn{1}{c|}{0.505(1)}           & 15.1          \\
\multicolumn{1}{c|}{2.311(1)}            & \multicolumn{1}{c|}{0.083(2)}          & \multicolumn{1}{c|}{0.871(4)}           & 6.0           \\
\multicolumn{1}{c|}{4.240(1)}            & \multicolumn{1}{c|}{0.056(2)}          & \multicolumn{1}{c|}{0.46(1)}            & 7.1           \\ \hline
\multicolumn{4}{c}{Sector~70}   \\ 
\hline
\multicolumn{1}{c|}{0.0953(4)}           & \multicolumn{1}{c|}{0.093(2)}          & \multicolumn{1}{c|}{0.514(3)}           & 5.0           \\
\multicolumn{1}{c|}{1.9429(3)}           & \multicolumn{1}{c|}{0.126(2)}          & \multicolumn{1}{c|}{0.287(2)}           & 5.0           \\
\multicolumn{1}{c|}{\textbf{2.06125(2)}} & \multicolumn{1}{c|}{\textbf{1.670(2)}} & \multicolumn{1}{c|}{\textbf{0.3773(1)}} & \textbf{52.9} \\
\multicolumn{1}{c|}{2.3090(4)}           & \multicolumn{1}{c|}{0.101(2)}          & \multicolumn{1}{c|}{0.876(2)}           & 4.9           \\
\multicolumn{1}{c|}{4.251(1)}            & \multicolumn{1}{c|}{0.048(2)}          & \multicolumn{1}{c|}{0.23(1)}            & 5.7           \\ \hline
\end{tabular}
\label{tab::TESS}
\end{table}
\renewcommand{\baselinestretch}{1}
\renewcommand{\baselinestretch}{0.75}
\setlength{\tabcolsep}{1.4pt}

\begin{table}[] 
\caption{Frequencies extracted for $\mu$\,Cet based on the combined photometric data of the  TESS sectors 42, 43 and 44} 
\medskip
\begin{tabular}{c|c|c|c}
\hline
\multicolumn{4}{c}{Sectors 42, 43, 44 } \\
\hline
\multicolumn{1}{c|}{Frequency,} & \multicolumn{1}{c|}{Amplitude,}   & \multicolumn{1}{c|}{\multirow{2}{*}{Phase}}               & \multirow{2}{*}{$S/N$}           \\ 
\multicolumn{1}{c|}{day$^{-1}$} & \multicolumn{1}{c|}{mmag}   & & \\
\hline
\multicolumn{1}{c|}{1.6494(2)}           & \multicolumn{1}{c|}{0.038(1)}          & \multicolumn{1}{c|}{0.389(5)}           & 4.5           \\
\multicolumn{1}{c|}{1.9364(1)}           & \multicolumn{1}{c|}{0.119(1)}          & \multicolumn{1}{c|}{0.855(2)}           & 7.4           \\
\multicolumn{1}{c|}{\textbf{2.06067(1)}} & \multicolumn{1}{c|}{\textbf{1.601(1)}} & \multicolumn{1}{c|}{\textbf{0.8775(1)}} & \textbf{56.1} \\
\multicolumn{1}{c|}{2.1620(1)}           & \multicolumn{1}{c|}{0.0814(1)}         & \multicolumn{1}{c|}{0.224(2)}           & 8.6           \\
\multicolumn{1}{c|}{2.3078(1)}           & \multicolumn{1}{c|}{0.095(1)}          & \multicolumn{1}{c|}{0.547(2)}           & 7.7           \\
\multicolumn{1}{c|}{4.1199(3)}           & \multicolumn{1}{c|}{0.028(1)}          & \multicolumn{1}{c|}{0.854(6)}           & 6.9           \\
\multicolumn{1}{c|}{4.2438(2)}           & \multicolumn{1}{c|}{0.047(1)}          & \multicolumn{1}{c|}{0.3400(4)}          & 10.1          \\ \hline
\end{tabular}
\label{tab::42_44}
\end{table} 
\renewcommand{\baselinestretch}{1}
\renewcommand{\baselinestretch}{0.75}
\setlength{\tabcolsep}{1.pt}
\begin{table}[t] 
\caption{Frequencies extracted for $\mu$\,Cet based on the combined photometric data from the TESS sectors 42, 43, 44 and 70}
\medskip
\begin{tabular}{c|c|c|c}
\hline
\multicolumn{4}{c}{Sectors~42, 43, 44, 70}       \\ 
\hline
\multicolumn{1}{c|}{Frequency,} & \multicolumn{1}{c|}{Amplitude,}   & \multicolumn{1}{c|}{\multirow{2}{*}{Phase}}               & \multirow{2}{*}{$S/N$}           \\ 
\multicolumn{1}{c|}{day$^{-1}$} & \multicolumn{1}{c|}{mmag}   & & \\
\hline
\multicolumn{1}{c|}{0.09803(1)}            & \multicolumn{1}{c|}{0.076(1)}          & \multicolumn{1}{c|}{0.846(2)}            & 5.3           \\
\multicolumn{1}{c|}{0.15938(1)}            & \multicolumn{1}{c|}{0.041(1)}          & \multicolumn{1}{c|}{0.731(3)}            & 4.3           \\
\multicolumn{1}{c|}{0.18302(1)}            & \multicolumn{1}{c|}{0.059(1)}          & \multicolumn{1}{c|}{0.183(2)}            & 4.3           \\
\multicolumn{1}{c|}{0.21797(1)}            & \multicolumn{1}{c|}{0.044(1)}          & \multicolumn{1}{c|}{0.862(3)}            & 4.3           \\
\multicolumn{1}{c|}{0.29545(1)}            & \multicolumn{1}{c|}{0.052(1)}          & \multicolumn{1}{c|}{0.505(2)}            & 4.6           \\
\multicolumn{1}{c|}{1.46026(2)}            & \multicolumn{1}{c|}{0.027(1)}          & \multicolumn{1}{c|}{0.818(5)}            & 4.5           \\
\multicolumn{1}{c|}{1.64930(2)}            & \multicolumn{1}{c|}{0.034(1)}          & \multicolumn{1}{c|}{0.591(4)}            & 4.9           \\
\multicolumn{1}{c|}{1.936831(4)}           & \multicolumn{1}{c|}{0.120(1)}          & \multicolumn{1}{c|}{0.835(1)}            & 7.0           \\
\multicolumn{1}{c|}{1.95582(2)}            & \multicolumn{1}{c|}{0.031(1)}          & \multicolumn{1}{c|}{0.470(3)}            & 4.6           \\
\multicolumn{1}{c|}{2.028954(6)}           & \multicolumn{1}{c|}{0.063(1)}          & \multicolumn{1}{c|}{0.548(2)}            & 6.7           \\
\multicolumn{1}{c|}{2.048764(6)}           & \multicolumn{1}{c|}{0.128(1)}          & \multicolumn{1}{c|}{0.373(1)}            & 9.2           \\
\multicolumn{1}{c|}{\textbf{2.0611761(4)}} & \multicolumn{1}{c|}{\textbf{1.530(1)}} & \multicolumn{1}{c|}{\textbf{0.62374(8)}} & \textbf{64.2} \\
\multicolumn{1}{c|}{2.071342(3)}           & \multicolumn{1}{c|}{0.198(1)}          & \multicolumn{1}{c|}{0.731(1)}            & 11.8          \\
\multicolumn{1}{c|}{2.08616(1)}            & \multicolumn{1}{c|}{0.047(1)}          & \multicolumn{1}{c|}{0.413(3)}            & 6.8           \\
\multicolumn{1}{c|}{2.12079(1)}            & \multicolumn{1}{c|}{0.042(1)}          & \multicolumn{1}{c|}{0.370(3)}            & 5.1           \\
\multicolumn{1}{c|}{2.149003(4)}           & \multicolumn{1}{c|}{0.152(1)}          & \multicolumn{1}{c|}{0.9842(8)}           & 8.5           \\
\multicolumn{1}{c|}{2.16250(1)}            & \multicolumn{1}{c|}{0.057(1)}          & \multicolumn{1}{c|}{0.996(2)}            & 7.4           \\
\multicolumn{1}{c|}{2.18326(2)}            & \multicolumn{1}{c|}{0.035(1)}          & \multicolumn{1}{c|}{0.721(4)}            & 6.0           \\
\multicolumn{1}{c|}{2.20901(2)}            & \multicolumn{1}{c|}{0.024(1)}          & \multicolumn{1}{c|}{0.742(5)}            & 4.4           \\
\multicolumn{1}{c|}{2.25848(2)}            & \multicolumn{1}{c|}{0.023(1)}          & \multicolumn{1}{c|}{0.19(1)}             & 4.3           \\
\multicolumn{1}{c|}{2.30812(1)}            & \multicolumn{1}{c|}{0.091(1)}          & \multicolumn{1}{c|}{0.667(1)}            & 7.5           \\
\multicolumn{1}{c|}{2.38224(2)}            & \multicolumn{1}{c|}{0.022(1)}          & \multicolumn{1}{c|}{0.12(1)}             & 4.7           \\
\multicolumn{1}{c|}{4.11961(2)}            & \multicolumn{1}{c|}{0.029(1)}          & \multicolumn{1}{c|}{0.613(4)}            & 8.0           \\
\multicolumn{1}{c|}{4.24348(1)}            & \multicolumn{1}{c|}{0.045(1)}          & \multicolumn{1}{c|}{0.166(3)}            & 11.2          \\ \hline
\end{tabular}
\label{tab::42_70}
\end{table}\renewcommand{\baselinestretch}{1}

The periodograms for the combined sectors 42--44 and 70 before and after the removal of the dominant frequency are shown in Figs~3a and 3b, respectively. Also in Fig.~3b, the gray line shows the power spectrum after extracting all the frequencies with $S/N > 4$.

\subsection{Archival Photometric Data}\label{Obsphotom}

To estimate the luminosity $L/L_{\odot}$ of $\mu$\,Cet, the radius $R^A$ and effective temperature $T^A_{\rm eff}$ of the primary component, we used the archival photometric data in various filters from the catalogs given in Table~4. The data were taken using the VizieR Photometry viewer service\footnote{\url{http://vizier.cds.unistra.fr/vizier/sed}}. For each filter, we selected one most reliable measurement.


The object parallax, according to the HIPPARCOS space mission, is \mbox{$\pi=38.71\pm1.31$~mas}, \mbox{$m_V=4\,.\!\!^{\rm m}26\pm 0\,.\!\!^{\rm m}01$} (ESA, 1997),  and according to the Gaia survey, it is \mbox{$\pi=37.59\pm0.24$~mas} (factor $RUWE=1.344$), \mbox{$m_G=4\,.\!\!^{\rm m}154\pm0\,.\!\!^{\rm m}003$} (Gaia Collab., 2022).  
 For the HIPPARCOS observations, the bolometric correction  $BC=-0\,.\!\!^{\rm m}11$ was taken from Allen's Astrophysical Quantities (Cox, 2000), for Gaia mission observations $BC=0\,.\!\!^{\rm m}08$---from the MESA bolometric correction tables\footnote{Modules for Experiments in Stellar Astrophysics: \url{https://waps.cfa.harvard.edu/MIST/}} (Dotter, 2016).

\begin{figure*}
\includegraphics[scale=0.6]{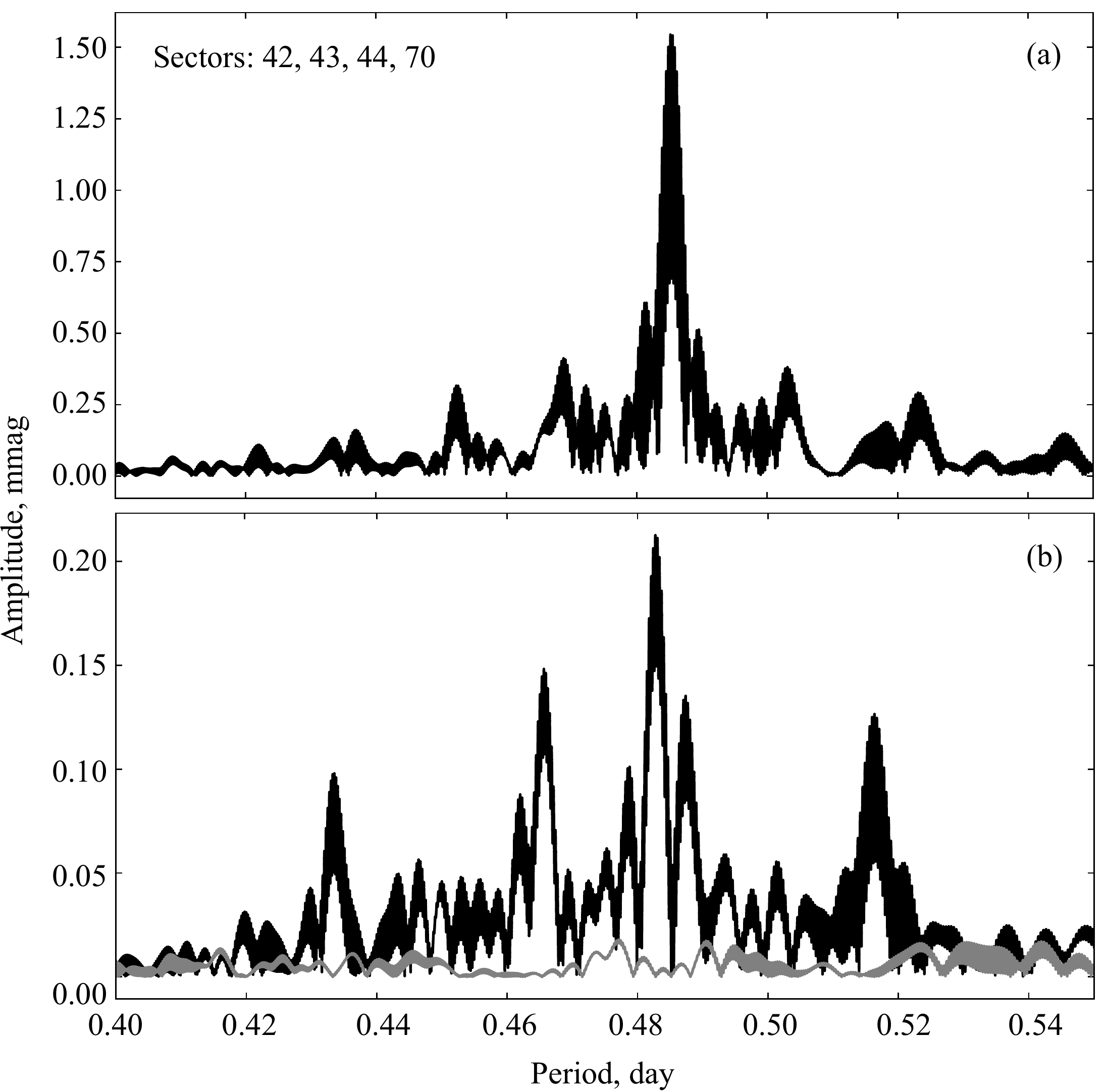}
\caption{Panel~(a): the periodograms of $\mu$\,Cet based on the combined data from sectors 42--44 and 70. Panel~(b): the same,
after removing the main frequency of 2.06~days$^{-1}$ and all frequencies with $S/N > 4$ (black and gray lines, respectively).}
\label{ps_2}
\end{figure*}

Ultraviolet observations in the region of up to 2000~\AA\ (observation date: September 8, 1993), also converted to absolute units, are taken from the IUE space observatory data archive\footnote{International Ultraviolet Explorer: \url{http://archive.stsci.edu/iue/}}.

\renewcommand{\baselinestretch}{0.75}
\setlength{\tabcolsep}{10pt}
\begin{table*}[]
		\caption{Selected photometric data from the Vizier portal}
 \medskip
 \label{tab:vizier}
	\begin{tabular}{l|l|l|l}
		\hline
		$\lambda$, \AA   & \multicolumn{1}{c|}{Filter}& \multicolumn{1}{c|}{Catalog } & \multicolumn{1}{c}{Ref.}\\
		\hline
		3530      & Johnson:\,$U$       				& II/7A/catalog         &  Morel and Magnenat (1978)  \\
		4020      & HIP:\,$Hp$          				& I/239/hip\_main       &  Perryman et al. (1997)  \\
		4203      & HIP:\,$BT$          				& I/239/hip\_main       &  Perryman et al. (1997) \\
		4442      & Johnson:\,$B$       				& II/7A/catalog         &  Morel and Magnenat (1978) \\
		5035      & Gaia/Gaia3:\,$ G_{\rm BP}$  & I/355/gaiadr3                 &  Gaia Collab. (2022)\\ 
		5318      & HIP:\,$VT$                          & I/239/hip\_main       &  Perryman et al. (1997) \\
		5537      & Johnson:\,$V$                       & II/7A/catalog         &   Morel and Magnenat (1978) \\
		5822      & Gaia/Gaia3:\,$G$                    & I/355/gaiadr3         &   Gaia Collab. (2022)\\
		6938      & Johnson:\,$R$                       & II/7A/catalog         &   Morel and Magnenat (1978)\\
		7620      & Gaia/Gaia3:\, $ G_{\rm RP}$   & I/355/gaiadr3               &  Gaia Collab. (2022)\\ 
        8779      & Johnson:\,$I$                       & II/7A/catalog         &   Morel and Magnenat (1978)\\
		12\,500   & Johnson:\,$J$                       & II/246/out            &  Cutri et al. (2003)  \\
		16\,300   & Johnson:\,$H$                       & II/246/out            &  Cutri et al. (2003) \\
		21\,900   & Johnson:\,$K$                       & II/246/out            &  Cutri et al. (2003) \\
		33\,500   & WISE:\,$W1$                         & II/311/wise           &  Cutri et al. (2012 \\
		34\,000   & Johnson:\,$L$                       & II/346/jsdc\_v2       &  Bourges et al. (2014) \\
		35\,499   & Spitzer/IRAC:\,3.6                  & J/AJ/163/45/table11   &  Rieke et al. (2022) \\                
		46\,000   & WISE:\,$W2$                         & II/311/wise           &  Cutri et al. (2012)  \\              
		50\,299   & Johnson:\,$M$                       & II/346/jsdc\_v2       &  Bourges et al. (2014 \\
		86\,100   & AKARI:\,$S9W$                       & II/297/irc            &  shihara et al. (2010) \\
		115\,598  & WISE:\,$W3$                         & II/311/wise           &  Cutri et al. (2012) \\
		115\,901  & IRAS:\,12                           & I/270/cpirss01        &   Hindsley and Harrington (1994)\\
		183\,898  & AKARI:\,$L18W$                      & J/MNRAS/471/770/table2 &  McDonald et al. (2017)\\
		220\,906  & WISE:\,$W4$                         & II/311/wise           &  Cutri et al. (2012) \\
		236\,746  & Spitzer/MIPS:\,24                   & J/ApJS/211/25/catalog &  Chen et al. (2014) \\
		238\,801  & IRAS:\,25                           & I/270/cpirss01        &  Hindsley and Harrington (1994) \\
		618\,497  & IRAS:\,60                           & I/270/cpirss01        &   Hindsley and Harrington (1994)\\
		714\,198  & Spitzer/MIPS:\,70                   & J/ApJS/211/25/catalog &   Chen et al. (2014) \\
	  1\,019\,492 & IRAS:\,100                        & I/270/cpirss01        &  Hindsley and Harrington (1994) \\ 
 \hline
	\end{tabular}
\end{table*}
\renewcommand{\baselinestretch}{1}

\section{RESULTS AND DISCUSSION}
\subsection{Orbit of the System}

\setlength{\tabcolsep}{2pt}
\begin{table}[t]
\caption{Positional parameters and brightness differences of $\mu$\,Cet based on the speckle interferometric measurements \medskip}
\label{tabl:simeas}
{
\begin{tabular}{l|r@{/}l|r@{\,$\pm$\,}l|r@{\,$\pm$\,}l|r@{\,$\pm$\,}l} \hline
\multicolumn{1}{c|}{Epoch}         & \multicolumn{2}{c|}{Filter}    & \multicolumn{2}{c|}{ $\rho$, mas}  & \multicolumn{2}{c|}{ $\theta$, deg}
& \multicolumn{2}{c}{ $\Delta m$, mag}                  \\
\hline
2018.9699 & 550&20   & 97.3&1           & 62.3&0.1       &      3.51&0.02   \\
2018.9699 & 694&10   & 94.6&1           & 61.7&0.1       &      3.11&0.02   \\
2019.0465 & 550&20   & 87.4&1           & 62.3&0.1       &      3.35&0.03   \\
2019.1508 & 550&20   & 66.8&1           & 58.9&0.1       &      3.33&0.03  \\
2019.2000 & 550&20   & 62.8&1           & 58.2&0.1       &      3.17&0.04   \\
2019.9393 & 550&20   & 148.5&1           & 237.1&0.1   &        3.47&0.03     \\
2019.9393 & 694&10   & 148.6&1           & 236.2&0.1   &        2.86&0.04     \\
2020.1774 & 550&20   & 176.2&1           & 235.9&0.1   &        3.37&0.02     \\
2020.6740 & 694&10   & 217.4&1           & 234.4&0.1   &        3.22&0.03     \\
2020.6740 & 800&100  & 218.6&1           & 234.3&0.1    &       3.17&0.02      \\
2020.7450 & 550&20   & 222.7&1           & 234.4&0.1      &     3.20&0.03    \\
2020.9031 & 550&20   & 232.3&1           & 234.0&0.1      &     3.77&0.02    \\
2021.6486 & 550&50   & 267.4&1           & 233.1&0.1      &     3.32&0.02    \\
2021.7360 & 550&50   & 271.4&1           & 232.9&0.1      &     3.83&0.02    \\
2021.9571 & 550&50   & 279.4&1           & 232.6&0.1      &     3.78&0.02    \\
2022.1131 & 550&50   & 282.8&1           & 232.5&0.1      &     3.65&0.03    \\
2022.943  & 550&50   & 302.7&1           & 231.7&0.1       &    3.49&0.03   \\
2023.1672 & 550&50   & 309.4&1           & 230.8&0.1       &    3.6?&0.04    \\
2023.7293 & 550&50   & 317&1           & 230.5&0.1       &      3.59&0.03    \\
\noalign{\smallskip} \tableline
\end{tabular}
}
\end{table}

\begin{figure*}
\includegraphics[width=0.82\textwidth]{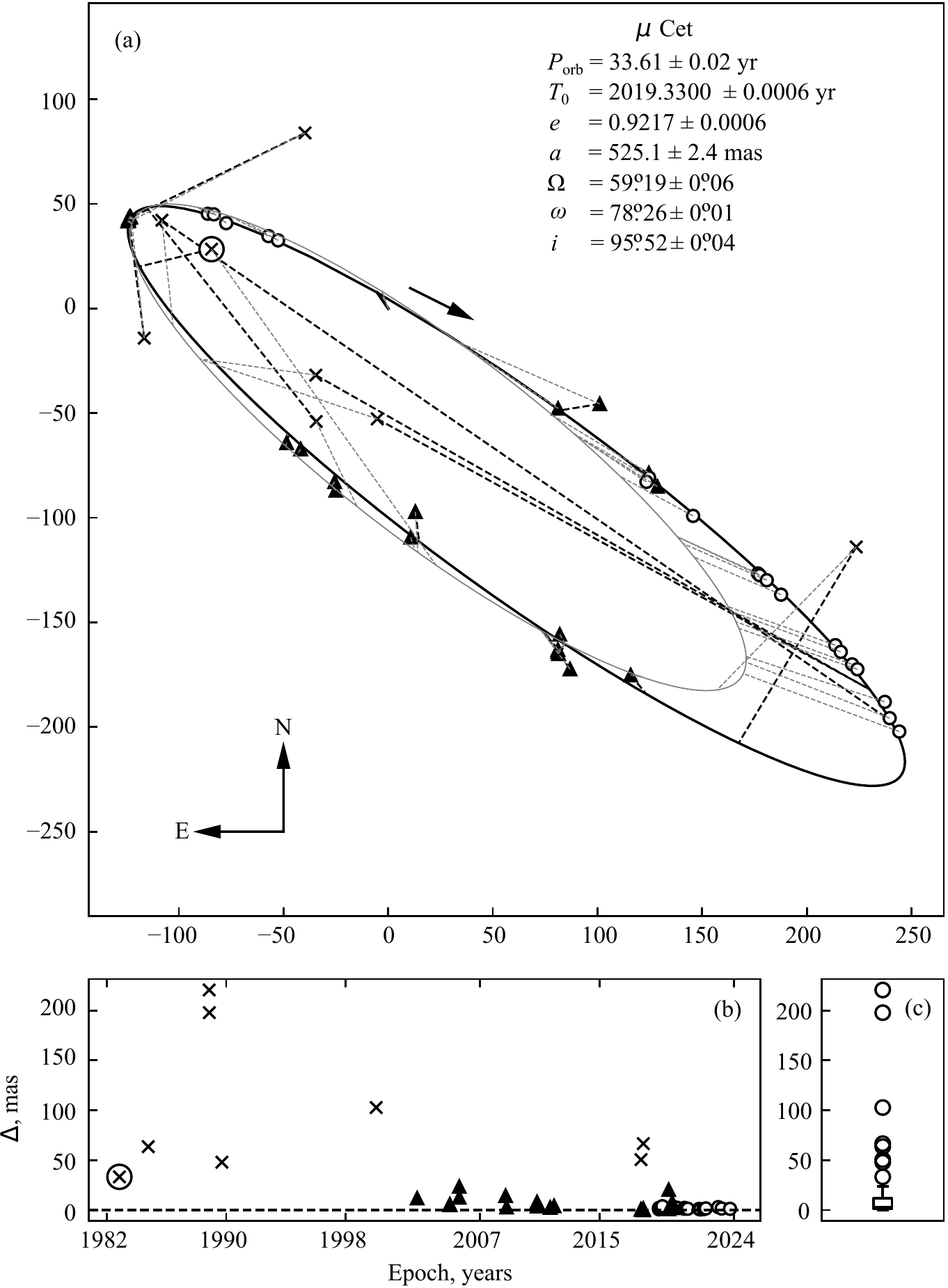}
\caption{Panel (a) gives the orbit of the $\mu$\,Cet pair. The crosses mark the points obtained in the \mbox{1980s} by speckle interferometry and lunar occultation data. The triangles mark the points from the literature, including the early BTA observations. The empty circles mark the data from the present work. The cross in the circle is the first resolved point (Tokovinin, 1985). Panel~(b) shows the residuals of individual points; panel~(c)---the corresponding span diagram.}
 \label{orbit}
\end{figure*}

The results of speckle interferometric measurements are presented in Table~5. To construct the orbit, we used the literature data from direct observations (see Dyachenko et al., 2019), new points obtained by Tokovinin et al. (2020) and the data from this paper. 
The orbit is shown in Fig.~4. The errors in calculating the orbital parameters are formal. A comparison of the orbital parameters of the preliminary solution and the one obtained in this paper are given in Table~6. 
Note that the high accuracy of determining the periastron passage point is associated with the small duty cycle
of the observations we have carried out near the time of passage and the large ($e = 0.9219$) eccentricity of the system.

Based on the results of new speckle interferometric observations, we did not detect a pair of weak components previously found from the observations in the IR range, using the lunar occultation method (Richichi et al., 2000). Note that their contribution is small for all the methods used in this work and does not affect the conclusions.

The data from the 1980s (Ismailov, 1992; Tokovinin, 1985) show significant deviations from the new orbital solution, which can be explained by the fact that the components are near the diffraction-limited resolution of the 1-m telescope of the Sanglok Observatory.

The resulting sums of the system masses are \mbox{$\Sigma M_{\rm hip} = 2.20 \pm 0.06\,M_{\odot}$} and \mbox{$\Sigma M_{\rm gaia} = 2.42 \pm 0.05\,M_{\odot}$} for the parallaxes of the HIPPARCOS and Gaia missions, respectively.
Using the corrected parallax ($\pi = 37.6847$~mas)  of the Gaia mission  
(Bailer-Jones et al., 2021)  \mbox{$\Sigma M_{\rm gaia} = 2.39 \pm 0.02\,M_{\odot}$}.

Based on the brightness difference measurements between the components at the 6-m telescope, the dependences for the MS stars
(Pecaut and Mamajek, 2013), and the status of the primary component of the system determined here, the secondary component is found to be a K0.5\,V dwarf.

\begin{table*}[]
\caption{Orbital parameters of $\mu$\,Cet} 
\label{orbitcompare} 
 \medskip
 \begin{tabular}{c|c|c|c|c|c|c|c}
  \hline
 $P_{\rm orb}$, yr & $T_{0}$, yr & $e$ & $a$, mas & $\Omega, $ deg &  $\omega, $ deg & $i,$ deg & Ref.$^*$ \\
\hline
    	25.4$\pm$0.3		&	2019.5$\pm$0.3	&	0.89$\pm$ 0.01	&	378$\pm$20	&	57.1$\pm$0.8	&	80.4$\pm$0.3	&	97.4$\pm$0.3	&	D2019	\\ 
     \hline
    	33.62$\pm$0.03		&	2019.3300$\pm$0.0006		&	0.9219$\pm$0.0006		&	525.9$\pm$2.6		&	59.18$\pm$0.06		&	78.26$\pm$0.01		&	95.5$\pm$0.04		& tw	\\ 
 \hline
\multicolumn{8}{p{16cm}}{*---Dyachenko et al. (2019); tw---the present paper.}
\end{tabular}
\end{table*}

\subsection{Magnetic Field}

Since the magnetic field, if strong enough, can affect the calculation of fluxes through the absorption variations in the line spectrum, the magnetic field of the A component was first estimated from both unpolarized and circularly polarized spectra. 
In the first case, the magnetic broadening of the Fe\,I~$\lambda$\,6336.82 line in the Elodie spectrum was studied as a function of $B_s$, the surface-averaged modulus of the $\mu$\,Cet magnetic field vector.
The synthetic spectrum was calculated using the {\tt Synmast} code (Kochukhov, 2007),  taking into account the contribution of the magnetic field to the formation of the spectral line. The {\tt Synmast} works in conjunction with the {\tt Binmag6} code (Kochukhov, 2018) for the visualization.
The specified line has an elevated  Lande factor $z=2.00$ and high intensity, and also lies in the long-wavelength part of the spectrum, thus increasing the Zeeman splitting. 
The atomic data, needed to calculate the line splitting pattern under the Zeeman effect were taken from the The Vienna Atomic Line Database (VALD)\footnote{\url{http://vald.inasan.ru}} (see Ryabchikova et al., 2015) using the ``{\tt Extract stellar}'' query in the {\tt long} format with the inclusion of the hyperfine structure (HFS splitting). To calculate the line profile, the $\mu$\,Cet\,A atmosphere models were used, with the parameters obtained in Section~3.3.2.

Polarimetric measurements of the surface-averaged longitudinal component of the magnetic field vector were performed using the BTA MSS spectra. The field was measured by two methods we use in the SAO RAS Laboratory for the Stellar Magnetism Study. The method of  magnetic field measurement was described in detail, e.g., in the papers by Semenko et al. (2014, 2022).

The MSS spectropolarimetry indicates that the visible longitudinal field does not exceed the determination error of about 100~G, which is consistent with the absence of additional magnetic broadening of the Fe\,I~$\lambda$\,6336.82 line. Therefore, to analyze the chemical composition of the atmosphere of $\mu$\,Cet\,A, we used the program for calculating the synthetic spectrum {\tt SynthVb} (Tsymbal, 1996) bundled with {\tt BinMag6} and independently the software package {\tt SME} (Valenti and Piskunov, 1996; Piskunov and Valenti, 2017), which do not take into account the magnetic field when forming the spectrum.

\subsection{Atmospheric Parameters and Stellar Radius}
\subsubsection{Photometry}\label{temp:R}

Blackwell and Shallis (1977) considered a method for determining $T_{\rm eff}$ and the angular diameter of a star $\theta$ from the photometric data (IRFM, infrared flux method). We calculate
\begin{equation}\label{eqTheta}
	\theta=2 \sqrt{\frac{F_{\lambda}^E}{F_{\lambda}^S}},
\end{equation}
where $F_{\lambda}^E$ is the observed flux on the Earth's surface, $F_{\lambda}^S$ is the continuum flux emerging from a unit area per unit of time from the surface of the star. 
This method is used for the observations in the infrared (1--13~$\mu$m) range of the spectrum due to a weak dependence  of $F_{\lambda}^S$ on $T_{\rm eff}$ and a small interstellar absorption in the specified wavelength range.
To calculate the $F_{\lambda}^S$  flux, we use the Kurucz grid of model atmospheres interpolated to the desired values of $T_{\rm eff}$, $\log g$, $[{\rm M}/{\rm H}]$.
The angular diameter error is calculated using the formula: $$\sigma_{\theta} = \sqrt{0.05^2+\sigma_{<\theta>}^2},$$ 
where the first term corresponds to a 10\%  error in the observed flux, and $\sigma_{<\theta>}$ is the error of weighted mean of the angular diameter, obtained for a set of infrared fluxes.

The effective temperature can be found by the formula: 
\begin{equation}\label{eqTeff}
	T_{\rm eff} =  \sqrt[4]{\dfrac{F^E}{(\theta/2)^2}},
\end{equation}
where $F^E$ is the integral flux registered on the Earth's surface.

Figure~5 shows the obtained values of angular diameter $\theta_A$ estimated from the IR fluxes. The contribution of the secondary component is subtracted from the final calculations.
The estimates of \mbox{$\theta_{\rm A }=  0.738 \pm 0.038$~mas}, \mbox{$R_{\rm A} =  2.07 \pm 0.11\,R_{\odot}$}
 were obtained from a set of measurements from 1 to 13~$\mu$m (in Fig.~5b this wavelength region and the  $\theta_{\rm A}$ error are indicated by a rectangle).
 Note that there is a significant error in the $L$ filter, and the data in $W2$ and $W3$, despite small errors, show a significant discrepancy, as do most of the measurements exceeding  13~$\mu$m. For these reasons, we excluded the measurements,  significantly differing from the majority  from the calculations of the mean $\theta_A$.
The estimate of $T^{\rm A}_{\rm eff}$ using formula~(2) yields the temperature $T=7380\pm180$~K.

\begin{figure*}
\includegraphics[scale=0.78]{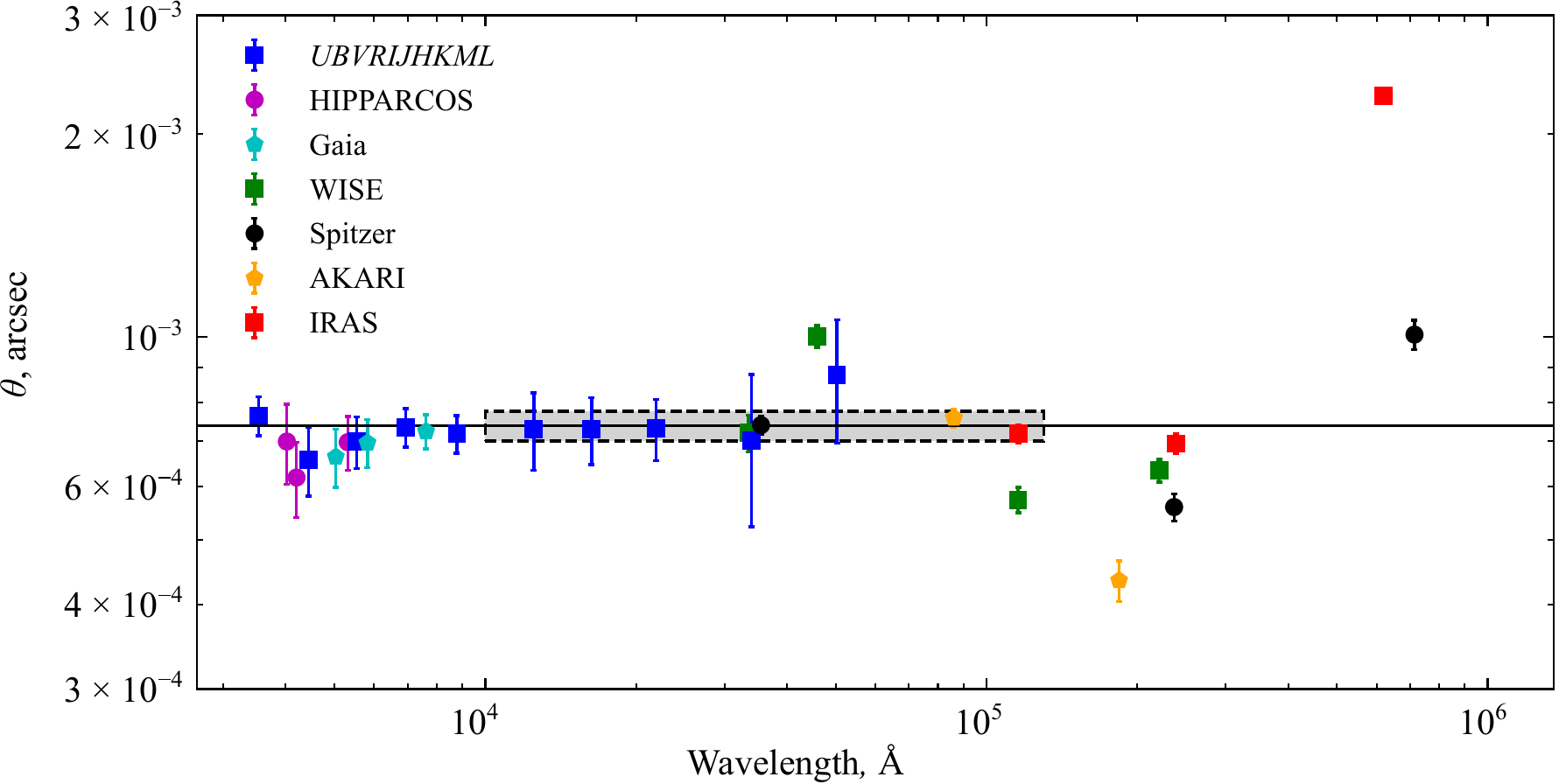}
	\caption{The angular diameter of the $\mu$\,Cet  A-component based on the multi-wavelength data. The estimate of $\theta_{\rm A}$ obtained in this paper is marked with a solid line; the rectangle shows the area of observations used and the 
  $\theta_{\rm A}$ error.}\label{thet}
\end{figure*}

\subsubsection{Spectroscopic analysis}\label{method:sp}

To find the parameters of the stellar atmosphere ($T_{\rm eff}$, $\log g$) and the abundances of some key chemical elements that determine the metallicity  $[{\rm M}/{\rm H}]$, we used the {\tt SME} (Spectroscopy Made Easy) software package written in the {\tt IDL} language for the automatic spectral analysis (Valenti and Piskunov, 1996; Piskunov and Valenti, 2017).

The atmospheric parameters, microturbulent velocity $\xi_t$, rotation velocity $v\,\sin i$ and radial velocity $v_r$ were determined by minimizing the $\chi^2$ criterion when approximating the selected sections of spectral observations with the synthetic spectra. A detailed study of the chemical composition in the LTE approximation was carried out both within the {\tt SME} package (using a group of lines of one element) and independently, using the {\tt SynthVb} code (Tsymbal, 1996) (individual lines).

In the {\tt SME}, one can choose to work with one of the three grids of model atmospheres: {\tt ATLAS9} (Castelli
and Kurucz, 2003), {\tt MARCS} (Gustafsson et al., 2008), {\tt LLmodels} (Shulyak et al., 2004). All the grids are calculated under the assumption of LTE and for a plane-parallel structure of the atmosphere.
We use the {\tt LLmodels} grid of models, since it takes into account the absorption coefficient in lines in more detail compared to the {\tt ATLAS9} models.

The atomic parameters of  spectral lines for calculating the synthetic spectrum were obtained from the VALD database using the  {\tt Extract stellar} query.

\begin{table*} \setlength{\tabcolsep}{4pt}
	\caption {A comparison of the $\mu$\,Cet atmospheric parameters, obtained from the spectra of Elodie and the MSS with the values from the literature} \label{tab:param2}
	\medskip
	\begin{tabular}{l|rrc|rrc|rrc|rr}
		\hline 		
		      \multirow{3}{*}{Parameter}& 
		\multicolumn{6}{c|}{Elodie} &  
	    \multicolumn{3}{c|}{MSS}   & 
	    \multicolumn{2}{c}{\multirow{2}{*}{ Gray et al.\,(2003)}}  \\ 
     \cline{2-7}
		 & 
        \multicolumn{3}{c|}{H$\alpha$\,+\,H$\beta$} &  
        \multicolumn{3}{c|}{H$\alpha$\,+\,H$\beta$\,+\,metals} &
        \multicolumn{3}{c|}{H$\alpha$\,+\,H$\beta$}   & 
        \multicolumn{2}{c}{}  \\ 	    
		\cline{2-12}
		             &               & $\sigma_{1}$  & $\sigma_{2}$ &               & $\sigma_{1}$  & $\sigma_{2}$ &               & $\sigma_{1}$ & $\sigma_{2}$&               & $\sigma$ \\ \hline
		$T_{\rm eff}$, К  & \textbf{7210} & 230           & 16           & \textbf{7260} & 360           & 20           & \textbf{7056} & 190          & 6           & \textbf{7225} & 115    \\
		$\log g$             & \textbf{3.8}  & 0.6           & 0.1          & \textbf{3.8}  & 1.7           & 0.2          & \textbf{3.5}  & 0.9          & 0.0         & \textbf{3.9}  &  \multicolumn{1}{r}{--~~}     \\
		$[{\rm M}/{\rm H}]$           & ${\bf -0.05}$& 0.16          & 0.02         & ${\bf-0.08}$& 0.22          & 0.05         &\multicolumn{1}{c}{--} &   \multicolumn{1}{c}{--}       &    \multicolumn{1}{c|}{--}     & \textbf{0.04} & 0.09  \\
		$\xi_t$, km\,s$^{-1}$     & \textbf{3.9}  & 1.0           & 0.1          & \textbf{3.8}  & 1.4           & 0.2          & \multicolumn{1}{c}{--}&\multicolumn{1}{c}{--} &\multicolumn{1}{c|}{--}& \textbf{3.2}  & \multicolumn{1}{r}{--~~~}\\
		$v\,\sin i$, km\,s$^{-1}$   & \textbf{50.3} & 9.2           & 0.7          & \textbf{49.3} & 12.3          & 1.0          &\multicolumn{1}{c}{--}& \multicolumn{1}{c}{--}&\multicolumn{1}{c|}{--}& \multicolumn{1}{c}{~--}& \multicolumn{1}{r}{--~~~}\\ \hline
	\end{tabular}
\end{table*}

In {\tt SME}, two errors are given:
$\sigma_{1}$ (cumulative), which is found from the cumulative distribution for the selected free parameter using all pixels inside the mask, and $\sigma_{2}$ (fitting error), which is an estimate of the efficiency of the algorithm implemented in {\tt SME}.
In the vast majority of cases, the uncertainty of the atmospheric parameter can be correctly taken as its cumulative error (for a detailed description, see Ryabchikova et al., 2016; Piskunov and Valenti, 2017).

\begin{figure*}
	\includegraphics[scale=0.28]{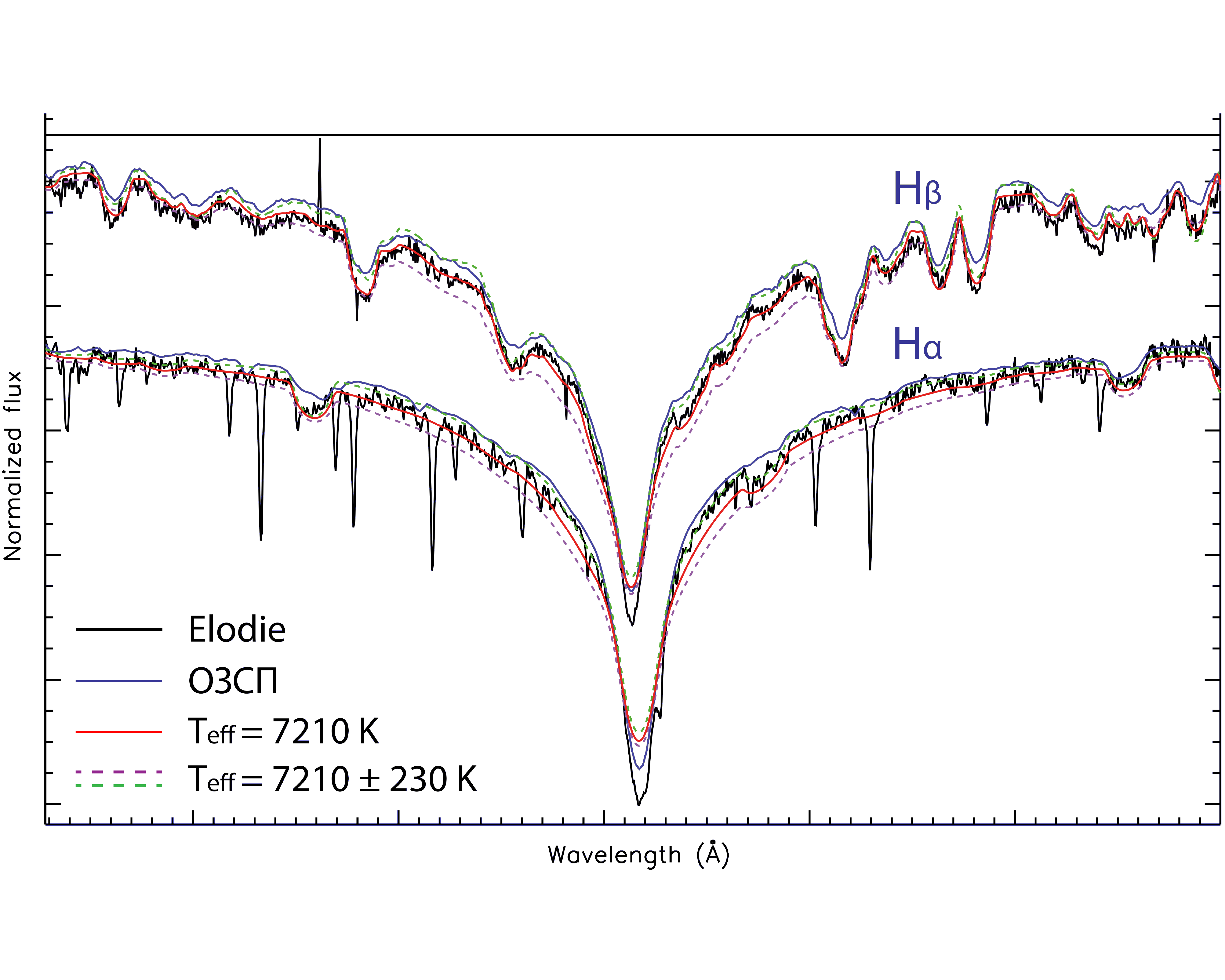}
	\caption{The H${\alpha}$ (bottom) and H${\beta}$ (top) line profiles. The black line describes the spectrum taken with the Elodie, the gray line (a narrower profile of the hydrogen line)---the spectrum with the MSS. The red solid line corresponds to the synthetic spectrum at $T_{\rm eff}=7210$~K. The synthetic spectra at the temperatures of $T_{\rm eff}=7210\pm230$~K are shown by the dashed lines.
	}\label{HbHa}
\end{figure*}

The choice of spectral regions for the work in the {\tt SME} (mask) is determined by the presence in these regions of spectral lines, sensitive to the variations of atmospheric parameters, for example, the lines of the Balmer hydrogen series (Ryabchikova et al., 2016).

Several spectral regions were selected in the Elodie spectrum: 4700--5000\,\AA\ and 6450--6650\,\AA,  containing the H${\beta}$ and H${\alpha}$ hydrogen lines, and the spectral ranges of 4400--4750\,\AA\ and 4950--6500\,\AA, containing metal lines. Due to the high degree of blending and uncertainty in the extraction of the spectral continuum, we do not use the spectral region shorter than 4400~\AA.

In the process of creating the spectral mask, telluric lines, lines with poorly known oscillator strengths, as well as the spectral regions with defects (dead pixels) and traces of cosmic particles, remaining after the primary spectrum processing, were excluded from consideration.
The hydrogen atom line cores (a region about $2.0$~\AA-wide near the line center) were also not used in the simulation.

The initial atmospheric parameters of $\mu$\,Cet in the {\tt SME}: \mbox{$T_{\rm eff}=7050\pm450$~K}, \mbox{$\log g=3.92\pm0.23$}, \mbox{$[{\rm M}/{\rm H}]=0.06\pm0.03$}, were estimated using the {\tt Templogg}~code (Kaiser, 2006) from the photometric indices in the Strömgren, Johnson, and Geneva photometry systems, taken from Hauck and Mermilliod (1998), Morel and Magnenat (1978), and Johnson et al. (1966), respectively.

\begin{figure*}
	\includegraphics[angle=0, scale=0.78]{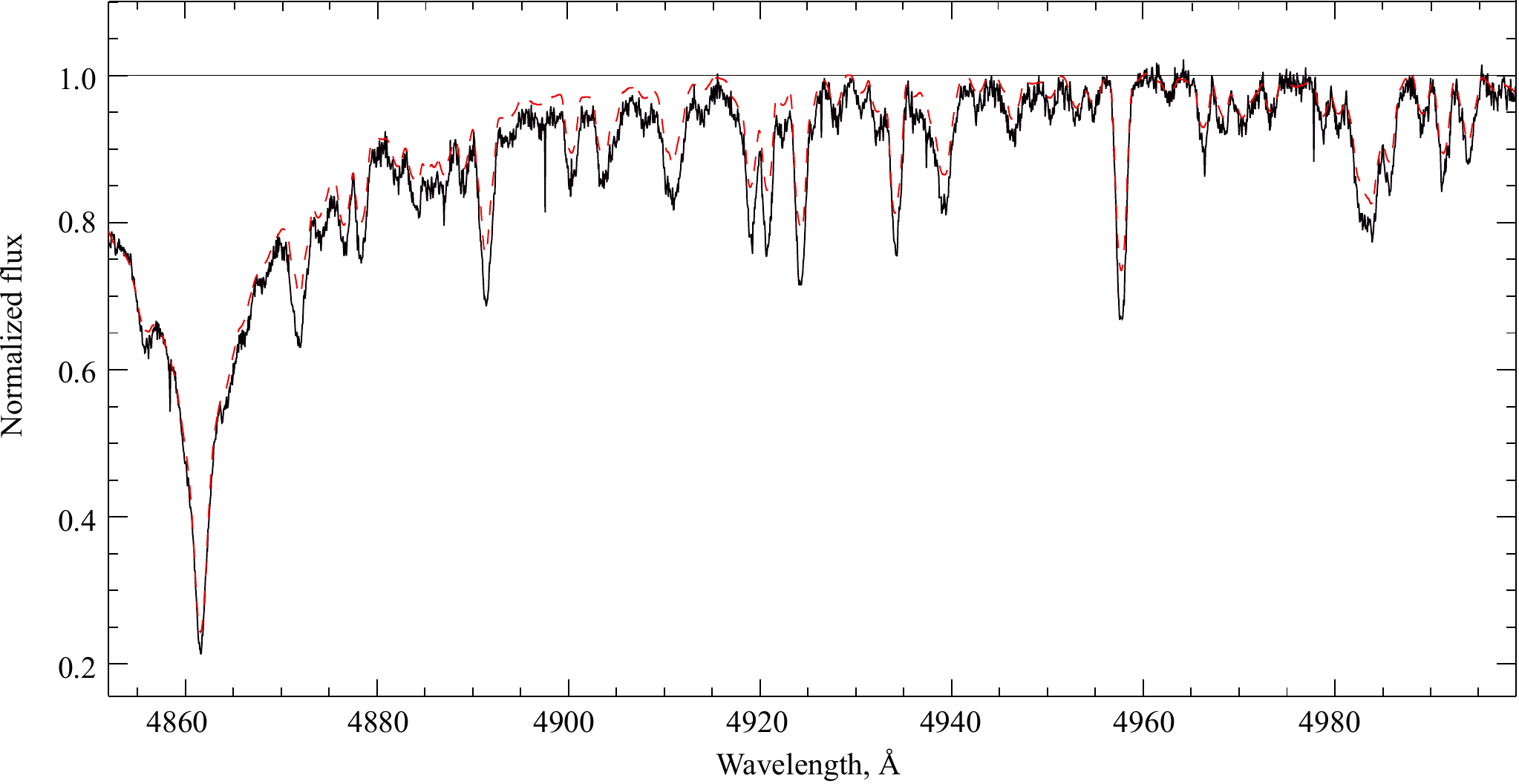}
	\caption{A comparison of the spectrum of $\mu$\,Cet from the Elodie (the solid line) with the spectrum of HD\,32115 (the dashed line).}\label{HD32115}
\end{figure*}

The results of determining the atmospheric parameters using the method described above are presented in Table~7. Due to the large value of $v\,\sin i$, the macroturbulent velocity in modeling the atmosphere is taken to be zero. The Table presents the parameters of the $\mu$\,Cet atmosphere obtained from the Elodie spectrum separately:
 \begin{list}{}{
\setlength\leftmargin{9mm} \setlength\topsep{0mm}
\setlength\parsep{0mm} \setlength\itemsep{1mm} }
\item[1)] from the regions with H${\alpha}$ and H${\beta}$ hydrogen line profiles, 
\item[2)] from the spectral orders 4400--4750~\AA\ and 4950--6500~\AA\ + H${\alpha}$ and H${\beta}$. 
 \end{list}
 The last two columns present the parameters obtained from the regions with H${\alpha}$ and H${\beta}$ of the MSS spectrum, as well as those, adopted from the literature sources.

The parameters of the atmosphere of $\mu$\,Cet we have obtained using the spectral analysis methods are in a very good agreement with the literature data. A comparison of the results of analysis of the hydrogen line profiles made independently from the spectra of two spectrographs indicates a lower $T_{\rm eff}$ found from the MSS spectrum (see the MSS data column in Table~7). 
This effect is partly explained by the known methodological complexity of subtracting the continuum level of the echelle spectra of early-type stars. It is possible that the continuum level in the region of hydrogen lines is subtracted more correctly in the spectrum, taken with the MSS (see Fig.~6). However, the average value of $T_{\rm eff} = 7380\pm180$~K from the photometry obtained with the Infrared Flux Method (IRFM) (see Section~3.3.1), supports a higher temperature.
The effective temperature of $7210\pm230$~K, determined from the hydrogen line profiles of the Elodie spectrum, corresponds to the temperature range  determined by different methods.

\begin{figure*}	\vspace{-3mm}
\includegraphics[scale=0.8]{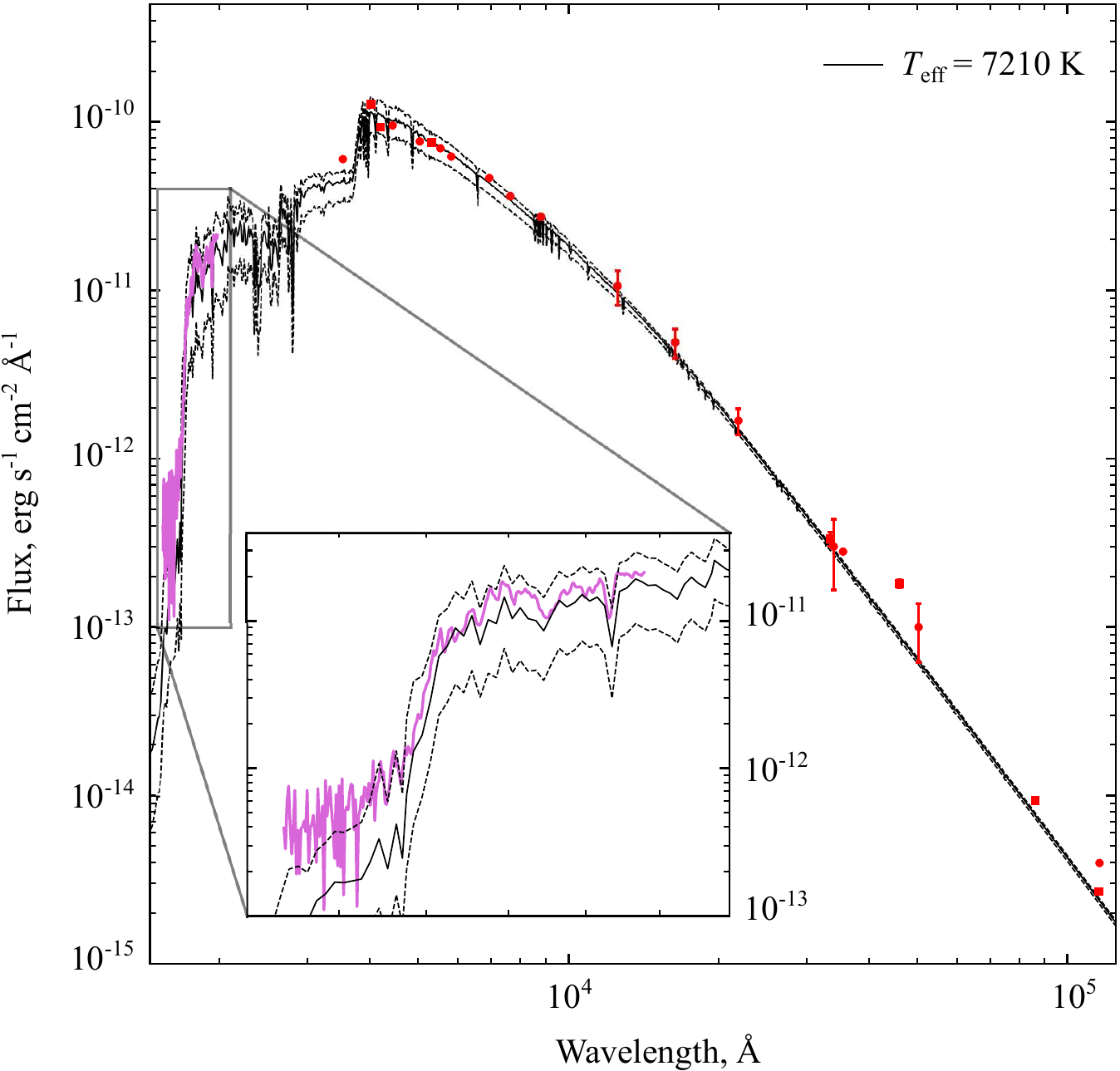}
\caption{A comparison of the observed and theoretical energy distributions of $\mu$\,Cet. 
The dots and the solid purple line demonstrate the photometric observations in individual filters and IUE, respectively. 
The theoretical energy distribution for the model with $T_{\rm eff}=7210$~K, $\log g=3.8$ is shown as a black solid line. The dashed lines correspond to the temperatures of $7210+230$~K, $7060-190$~K.} \label{SED}
\end{figure*}

We have confirmed our $\mu$\,Cet\,A effective temperature measurement and, partially, the metallicity by a comparison with the spectrum of a normal star HD\,32115 with the atmospheric parameters  7250/4.2/0.0, \mbox{$\xi_t=2.3$~km\,s$^{-1}$} (see Mashonkina et al., 2020), shown in Fig.~7. The main indicators of $T_{\rm eff}$, the hydrogen line profiles, coincide. 
Although all the metal lines in the spectrum of $\mu$\,Cet are deeper, this is easily explained by a greater value of microturbulent velocity in the atmosphere of $\mu$\,Cet (3.9~km\,s$^{-1}$).

\subsubsection{Spectral Energy Distribution}

Additionally, we have checked the atmospheric parameters by comparing the observed and model spectral energy distributions. 
In Fig.~8, the dots show the photometric measurements taken from open data sources. 
The purple solid curve shows the spectral energy distribution obtained at the IUE\footnote{International Ultraviolet Explorer {\url {https://sci.esa.int/web/iue}}}.
The black solid curve shows the model distribution of the $\mu$\,Cet flux, calculated for the parameters of the ``hot'' (7210~K) $\mu$\,Cet atmosphere from Table~7. The dotted lines indicate the distributions calculated for the maximum range of variations of both effective temperatures, taking into account their errors $1\,\sigma_{1}$. In all the theoretical flux calculations, the stellar radius \mbox{$R_A= 2.07\pm0.11\,R_{\odot}$} was used.
 
Both models of the atmosphere of the $\mu$\,Cet\,A component agree with the observations, within $\sigma_1$ for $T_{\rm eff}$ and within $\sigma$ for $R_A$. 

The magnitude difference between the components of the $\mu$\,Cet system in three filters  \mbox{$\Delta m_{500} =3\,.\!\!^{\rm m}5\pm 0\,.\!\!^{\rm m}3$}, \mbox{$\Delta m_{694} =3\,.\!\!^{\rm m}1\pm 0\,.\!\!^{\rm m}3,$} \mbox{$\Delta m_{800} =3\,.\!\!^{\rm m}2$} 
has been converted into the flux ratios. By interpolating the photometric points, we have obtained the fluxes at the wavelengths of 500, 694 and 800 nm. By solving two equations with two unknowns at each of the three wavelengths, we obtained the fluxes for the components A and B.
We estimate the parameters of component B as \mbox{$T_{\rm eff}^B=5250$~K}, corresponding to the MS stars by \mbox{$m_V=3\,.\!\!^{\rm m}5$}
 fainter than the main component (see~Section~3.6), and $R_B=0.80\pm0.15~R_{\odot}$. The final contribution of the component B to the observed spectrum is small enough to be neglected when estimating the chemical abundance.

\subsection{Chemical Composition}
The MSS spectropolarimetry indicates that the apparent longitudinal field does not exceed the determination error, which is about 100~G. 
Hence, to calculate the chemical abundance, we used the {\tt SynthVb} code along with the {\tt BinMag6} and independently the {\tt SME} software package, which does not take into account magnetic line broadening. 
The chemical abundance analysis was carried out for two sets of atmospheric parameters obtained from the hydrogen lines.

\setlength{\tabcolsep}{4.5pt}
\begin{table*}
	\caption{Abundances of 14 chemical elements in the Sun and $\mu$\,Cet. The calculations are shown for the ``hot'' and ``cold'' atmosphere parameters from Table~7. For the elements with an asterisk *, the abundances are obtained within the {\tt Binmag6} and {\tt SynthVb} are given, for the remaining elements, the abundances calculated within {\tt SME} are listed\medskip}\label{tab::Abun}
	\begin{tabular}{c|c|c|c|c|r|c|c|r|r}
		\hline
		\multirow{2}{*}{Element}& \multicolumn{2}{c|}{Sun}                &    \multicolumn{3}{c|}{$T_{\rm eff}=7210$~K}                   & \multicolumn{3}{c|}{$T_{\rm eff}=7060$~K}                             & \multirow{2}{*}{$\Delta_{\rm NLTE}$}                \\ \cline{2-9}
		 & $\log\varepsilon_{\bigodot}$ &\multicolumn{1}{c|}{$N$}  & $\log\varepsilon$ & $\sigma$& $[N_{\rm el}/N_{\rm tot}]$         & $\log\varepsilon$ & $\sigma$ & $[N_{\rm el}/N_{\rm tot}]$ &  \\ \hline
		C*    & 8.46               & 4   & 8.54 & 0.08                 & 0.08~~      & 8.45               & 0.13     & $-0.01$~~     & $-0.10$~~           \\
		Mg   & 7.55               & 8   & 7.65 & 0.22                 & 0.10~~      & 7.70               & 0.23     & 0.15~~      & $-0.03$~~           \\
		Si   & 7.51               & 4   & 7.70 & 0.29                 & 0.19~~      & 7.68               & 0.29     & 0.17~~      & $-0.07$~~           \\
		Ca   & 6.30               &--   & 6.45 & 0.18                 & 0.15~~      & 6.42               & 0.19     & 0.12~~      & $-0.14$~~           \\
		Sc   & 3.14               &--   & 2.95 & 0.21                 &    $-$0.19~~  & 2.85               & 0.25    & $-0.29$~~      & \multicolumn{1}{c}{--}              \\
		Ti   & 4.97               &--   & 4.87 & 0.23                 & $-$0.10~~     & 4.72               & 0.22    & $-0.25$~~     & 0.03~~            \\
		Cr   & 5.62               &--   & 5.63 & 0.29                 & 0.01~~      & 5.52               & 0.29     & $-0.10$~~      & \multicolumn{1}{c}{--}              \\
		Fe   & 7.46               &--   & 7.37 & 0.19                 & $-$0.09~~     & 7.28               & 0.19     & $-0.18$~~      & $-0.01$~~           \\
		Ni   & 6.20               &--   & 6.09 & 0.41                 & $-$0.11~~      & 6.01               & 0.42     & $-0.19$~~      & \multicolumn{1}{c}{--}              \\
		Zn   & 4.56               &--   & 4.01 & 0.25                 & $-$0.55~~     & 4.00               & 0.25     & $-0.56$~~     & 0.13~~            \\
		Sr   & 2.83               & 2   & 3.27 & 0.34                 & 0.44~~      & 3.15               & 0.32    & 0.32~~      & $-0.05$~~           \\
		Y    & 2.21               &--   & 2.14 & 0.35                 &  0.03~~     & 1.95               & 0.40     & $-0.26$~~      &\multicolumn{1}{c}{--}             \\
		Ba*   & 2.27               & 4   & 2.24 & 0.15                 & $-$0.03~~     & 2.14               & 0.26     & $-0.13$~~     & $-0.17$~~           \\
		Eu   & 0.52               & 2   & 0.73 & 0.28                 & 0.21~~      & 0.58               & 0.27     & 0.06~~      & \multicolumn{1}{c}{--}              \\ \hline
	\end{tabular}
\end{table*}
\begin{figure*}
\includegraphics[width=\textwidth]{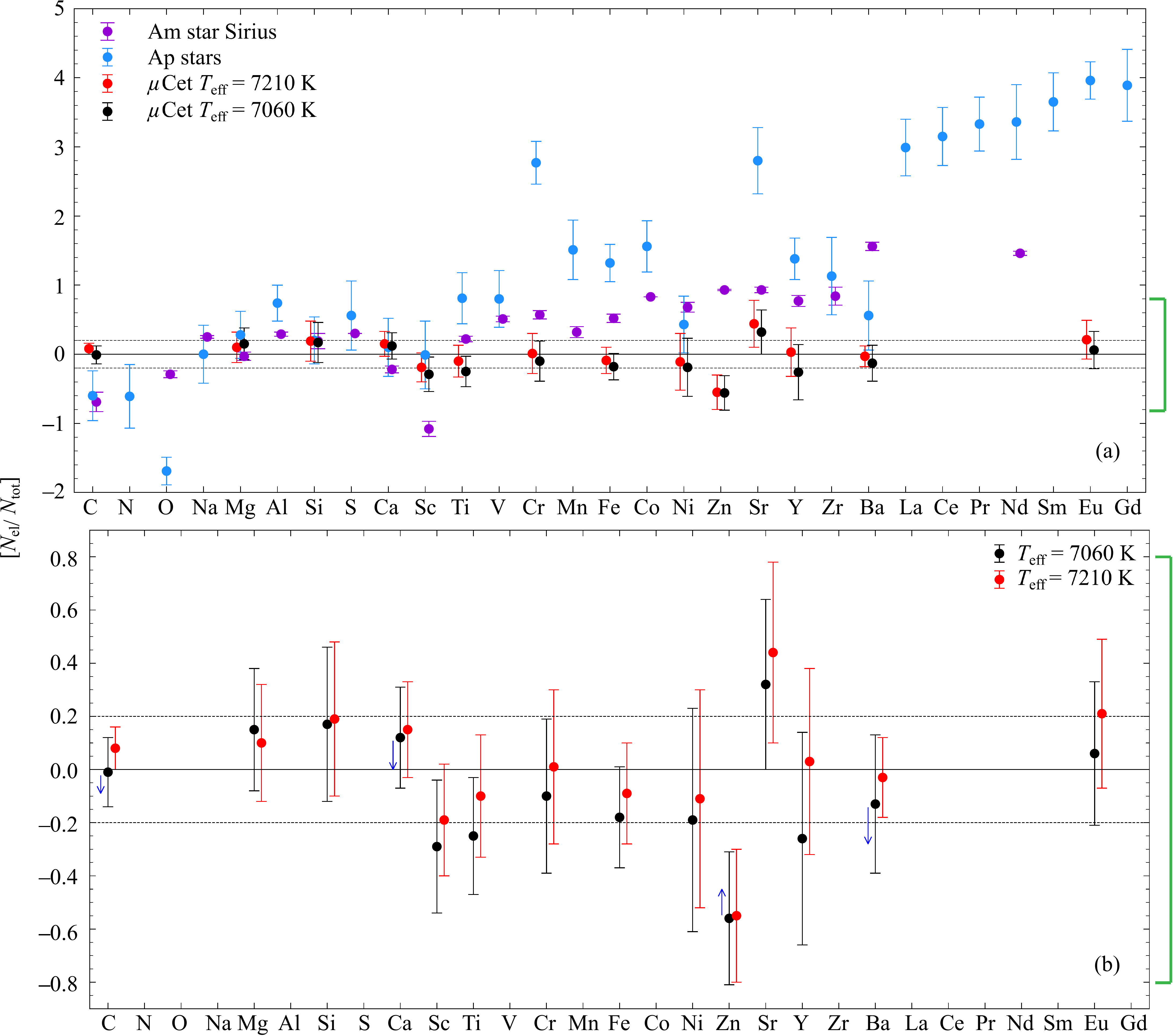}
\caption{Panel (a): a comparison of the elemental abundances in the atmosphere of $\mu$\,Cet\,A with the abundances of the same elements in the Am star Sirius and in Ap stars. Panel (b): elemental abundances in the atmosphere of $\mu$\,Cet\,A relative to the Sun for two model atmospheres. The characteristic uncertainty of  $\pm 0.2$~dex is indicated by the dashed lines.
The arrows show the approximate inclusion of non-LTE effects in the elemental abundances.}\label{Abun2}
\end{figure*}

We determined the abundances of  such elements as Ca, Sc, Ti, Cr, Fe, Ni, Zn, Y in the {\tt SME} package from the Elodie spectral regions that do not contain hydrogen line profiles. 
For Mg, Si, Sr, Eu, the abundances were determined in {\tt SME} with masks in which only the observed lines of these atoms were isolated.
Such a choice of mask is especially important for the peculiar elements Sr, Eu, since the lines of these atoms are few and weak. 
For the elements C and Ba, the abundances were obtained by averaging the abundances calculated separately for each line using the {\tt SynthVb} and {\tt Binmag6}. The average abundance of an element $\log\varepsilon$, the abundance of an element relative to the Sun $[N_{\rm el}/N_{\rm tot}]$, the abundance error $\sigma$ (only for C, Ba, for other elements the cumulative error from {\tt SME} is taken as $\sigma$) of an element were calculated using the formulas:

\begin{eqnarray}
	\log\varepsilon=\log(N_{\rm el}/N_{\rm tot})+12.04,\nonumber\\
	\left[ N_{\rm el}/N_{\rm tot} \right] = \log\varepsilon-\log\varepsilon_{\odot},\nonumber\\
	\sigma = \sqrt{\frac{\sum_{i=1}^N (\log\varepsilon_i-\log\varepsilon)^2}{N-1}},\nonumber
\end{eqnarray}
where $N_{\rm el}$~ is the concentration of atoms of a given element in the atmosphere, $N_{\rm tot}$~is the concentration of all atoms; $\log\varepsilon_{\odot}$ is the abundance of an element in the solar atmosphere, in the system where $\log\varepsilon({\rm H})=12.00$, $\log\varepsilon({\rm He})=11.04$; $\log\varepsilon_{i}$~ is the abundance of an element determined from the $i$-th line.
Solar abundances of elements are taken from Asplund et al. (2021). 

The results of finding the chemical element abundances in the atmosphere $\mu$\,Cet are given in Table~8.
If the abundance of an element is determined using the {\tt SynthVb} code or using {\tt SME} with special masks in which only the observed lines of these atoms/ions were selected, then the table shows the number of lines of a specific atom/ion, used in the calculation.

To estimate the approximate non-LTE corrections of $\Delta_{\rm NLTE}$ to the abundance relative to the Sun in the LTE approximation $[N_{\rm el}/N_{\rm tot}]$, we used the data for the star HD\,32115 from Mashonkina et al. (2020). The correction for Zn is taken from Sitnova et al. (2022).

The $[N_{\rm el}/N_{\rm tot}]$  abundances in the $\mu$\,Cet\,A atmosphere are shown in Fig.~9b. The correction values for the elements C, Ca, Zn, Ba are marked with the arrows. Panel (a) compares the chemical composition of $\mu$\,Cet\,A with the chemical composition of the atmosphere of the Am star Sirius (Mashonkina et al., 2020), as well as with the averaged chemical composition of nine Ap stars from Romanovskaya (2023).

Within $\pm0.2$~dex, the chemical composition of the $\mu$\,Cet\,A atmosphere is close to solar without significant systematic deviations.

\subsection{Pulsations}

\begin{figure*}		
\includegraphics[scale=0.75]{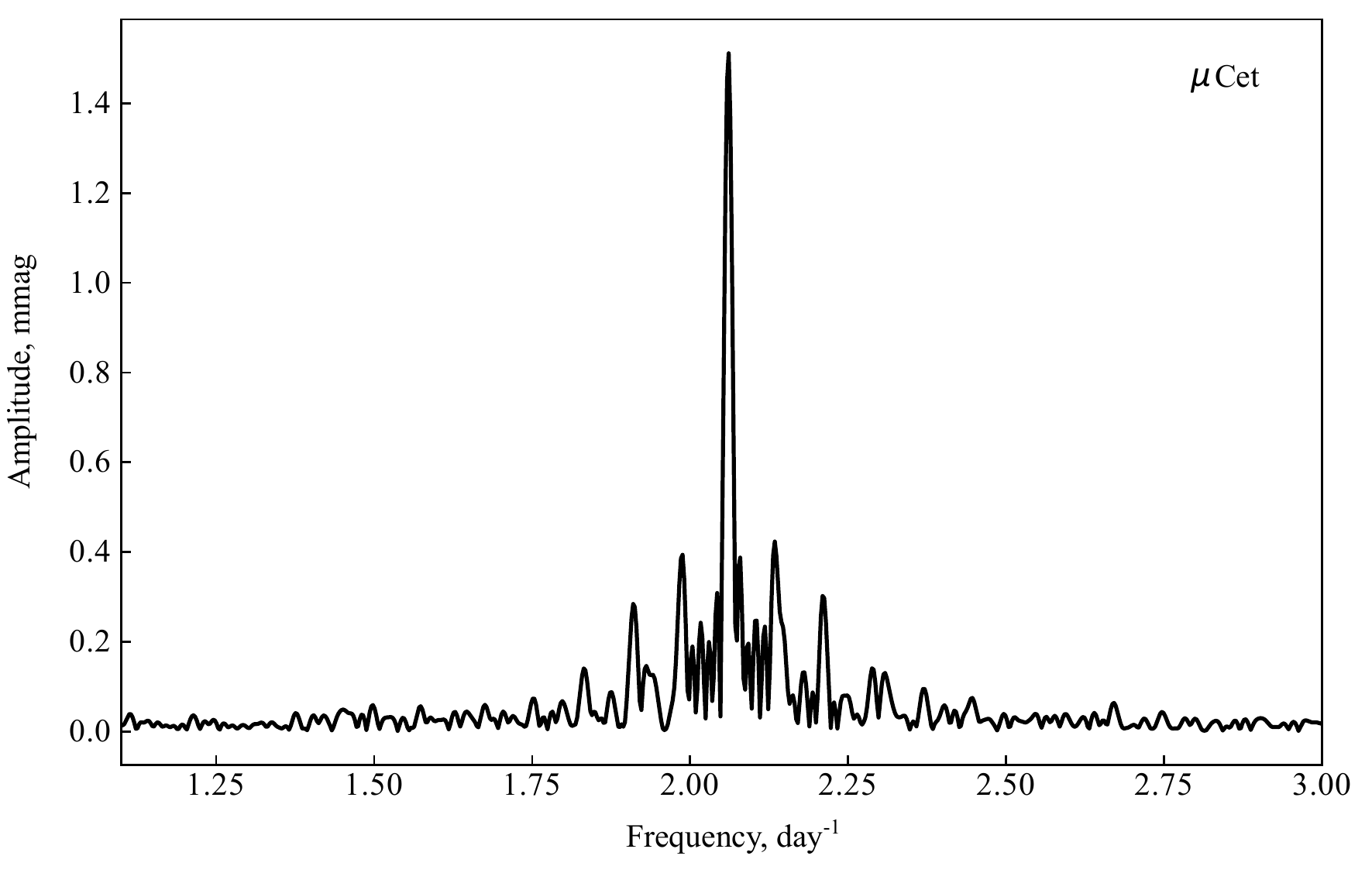}
\caption{The Lomb--Scargle periodogram of $\mu$\,Cet based on the TESS mission data.}\label{tessps}
\end{figure*}

\begin{figure*}	
\includegraphics[scale=0.59]{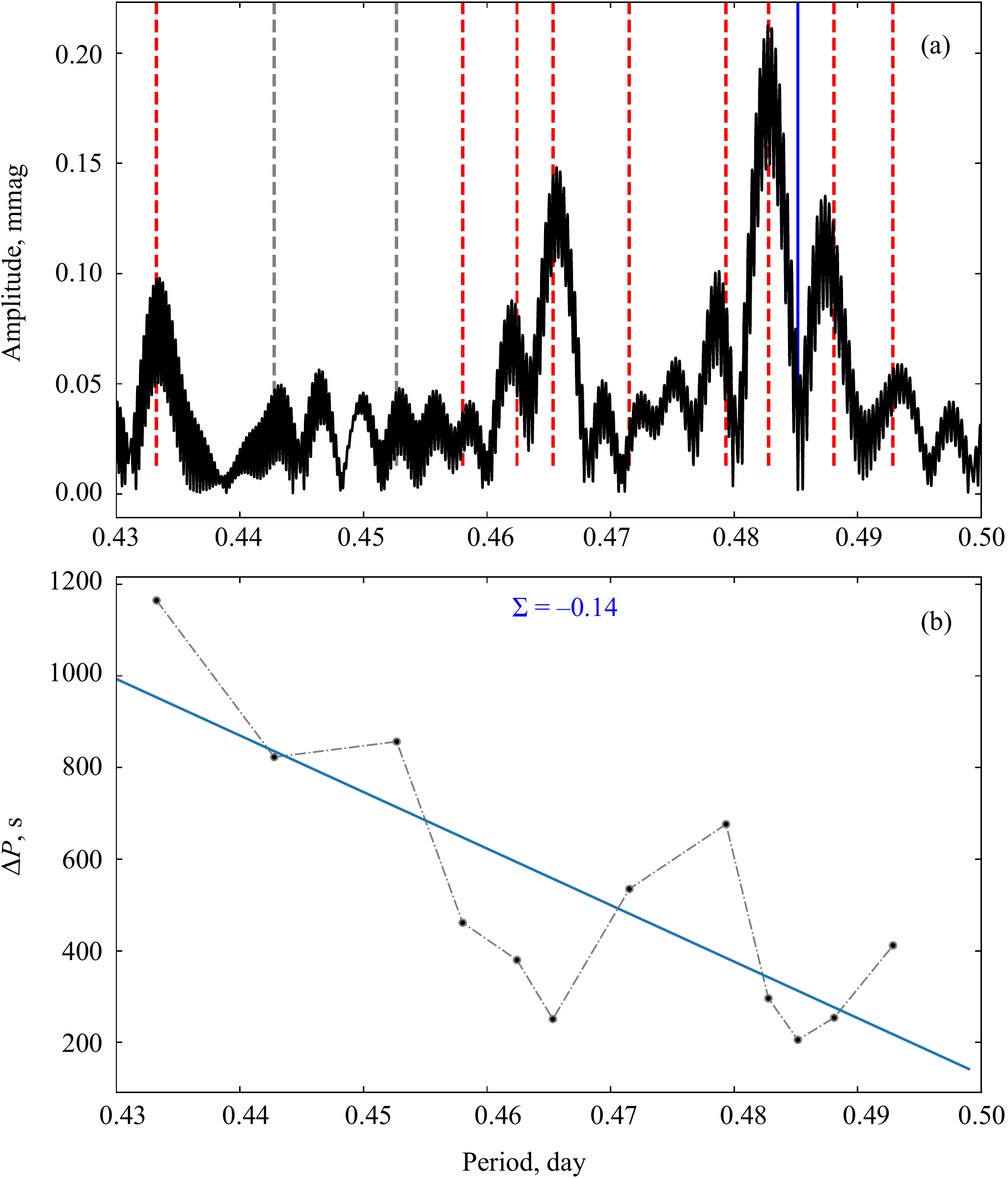}
	\caption{Panel (a)~ shows a periodogram of $\mu$\,Cet with the main frequency removed (the position is marked by the blue solid line). Panel (b)~  gives the ``$\Delta P$\,--\,$P$'' dependence, where $\Delta P$ is the difference between successive periods; the blue line is the approximating straight line with the slope  $\Sigma = -0.14$.}
	\label{dP-P}
\end{figure*}

According to the photometric data of the TESS mission, the main peak with a half-amplitude of $1.51\pm0.05$~mmag (???) corresponds to a period of $0.485\pm0.002$~days. The Lomb-Scargle periodogram with the corresponding profile is shown in Fig.~10. The indicated value of the period corresponds to
$S/N = 28$, which is calculated by estimating the background noise. The variability period is not typical for the  $\delta$\,Sct-type stars. The obtained  period and a short amplitude correspond to the $\gamma$\,Dor-type variables, which is consistent with the spectral class of the main component of $\mu$\,Cet.

Assuming that the values in the Johnson $V$ filter \mbox{$m_{V} = 4\,.\!\!^{\rm m}26$} are close to the values of the 550/20 and 550/50~nm speckle interferometer filters, we obtain in the $V$ band for the primary and secondary components \mbox{$m_V^{{\rm A}} = 4\,.\!\!^{\rm m}30$}, \mbox{$m_V^{\rm {B}}= 7\,.\!\!^{\rm m}80$}, respectively. Such a large difference in the component brightness allows us to neglect the influence of the secondary component on the photometric variability of the system, as well as on its spectrum.

The variability of $\mu$\,Cet may be due to pulsations of the primary component in the gravitational modes, typical of $\gamma$\,Doradus-type objects ($\gamma$\,Dor)---the MS stars of early F or late A spectral classes and masses of \mbox{$1.4\,M_{\odot}$ $\le M \le 2.0\,M_{\odot}$} (Kaye et al., 1999). The pulsations of these objects are multiperiodic with periods ranging from 0.3 to 4~days.

Due to their thin convective envelopes (zones???), $\gamma$\,Dor stars do not possess significant magnetic fields (Schatzman, 1962), as is the case with $\mu$\,Cet. The emergence of large spots and, as a consequence, a brightness variability due to rotation, are unlikely. 
Based on the location of variable stars on the Hertzsprung--Russell diagram, one can make an assumption about the nature of the variability (see Fig.~7 in Fetherolf et al., 2023):
$\mu$\,Cet is located in the region of A--F-class objects with short periods (less than one day), a large proportion of which are pulsating stars.

The ``$\Delta P$\,--\,$P$'' dependence is the main tool for measuring the circumnuclear rotation velocity, which is the subject of many studies (Van Reeth et al., 2016; Christophe et al., 2018a; Van Reeth et al., 2018; Li et al., 2019, 2020; Takata et al., 2020; Pedersen et al., 2021). Despite the scarcity of available data, we constructed the dependence of the difference between the successive periods $\Delta P$ on the period value ($P$) for $\mu$\,Cet (Fig.~11).
We may with care speak about the downward trend, that is, about a decrease in the distance between the periods. 
In turn, this indicates the prograde pulsation modes (pulsations propagate along the rotation). 
The slope ($\Sigma$) of the ``$\Delta P$\,--\,$P$'' dependence serves as an indicator of the stellar rotation velocity. 
We approximated the obtained dependence with a linear function and obtained $\Sigma = -0.14$, while the typical values for the prograde dipole modes are approximately $-0.03$ (Ouazzani et al., 2016).
A small amount of data and gaps in the general photometric series reduce the $S/N$ ratio, so we consider the obtained 
\mbox{``$\Delta P$\,--\,$P$''} dependence to be preliminary.

\subsection{Position on the H-R Diagram}\label{AgeMass}
\begin{figure*}
\includegraphics[scale=0.9]{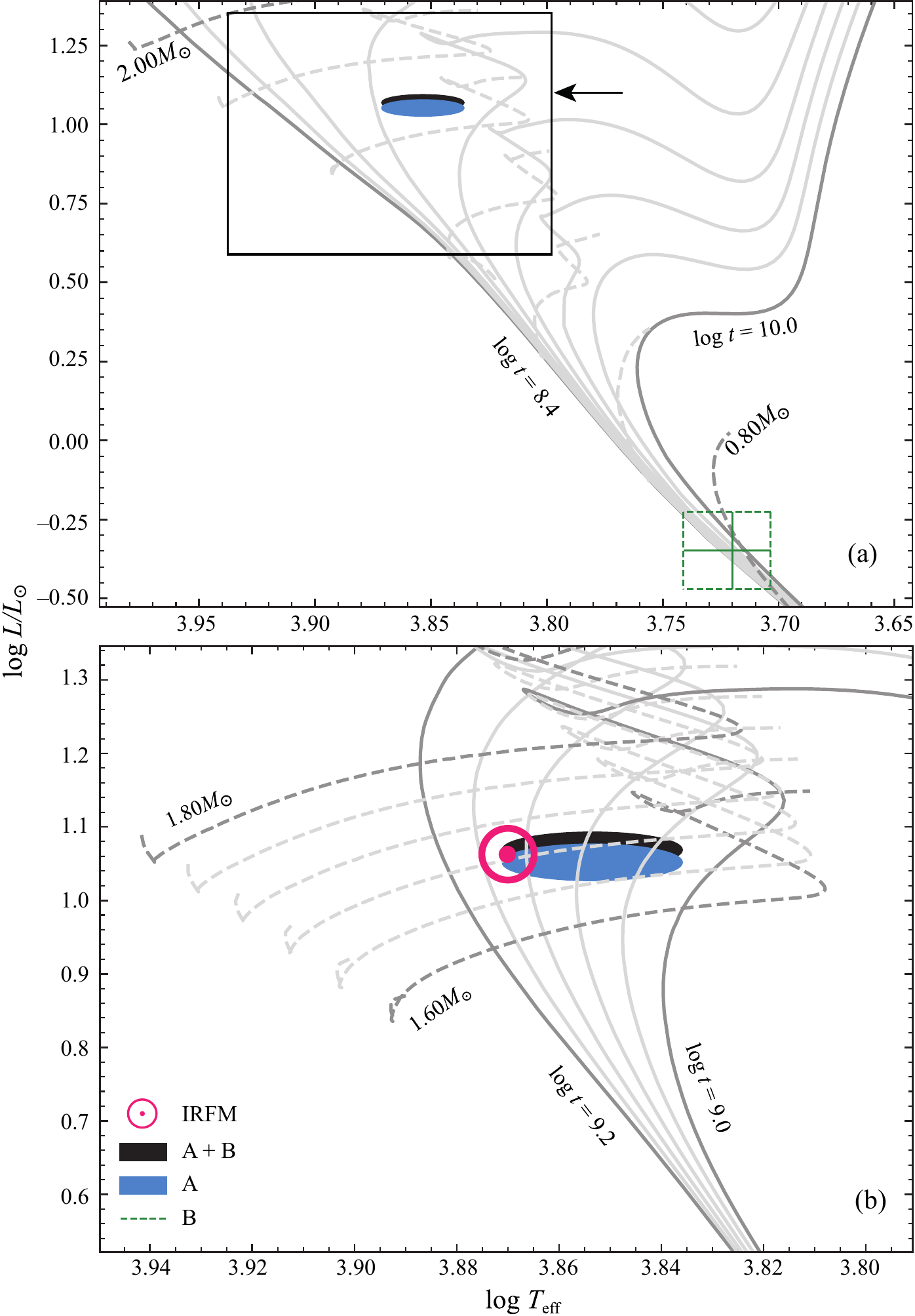}
 \caption{The position of $\mu$\,Cet on the H-R diagram (a black ellipse), the positions of components A (a blue ellipse) and B (an area highlighted by green lines). The radius of component~A is marked in blue. A part of the figure in panel (a) is shown in an enlarged scale (indicated by an arrow) in panel (b).}\label{HR}
\end{figure*}

We can estimate the mass and age of the object under study based on the atmospheric model and the chemical composition  analysis. 
Figure~12a,\,b demonstrates: the position of $\mu$\,Cet on the theoretical Hertzsprung--Russell diagram (a black ellipse) in the Gaia and HIPPARCOS observations (see Section~2.4), for ``cold'' and ``hot'' model atmospheres; 
a set of isochrones and evolutionary tracks taken from the MIST (MESA Isochrones and
Stellar Tracks, Dotter, 2016) given the solar chemical composition of the atmosphere and the rotation velocity of the star less than half of the critical one.

Since the luminosity ratio \mbox{$L_A/L_B \simeq 10^{0.4\times\Delta m_{500}}$} and the sum \mbox{$L_A+L_B = L$} are known, 
we have shown in Fig.~12 the positions of A (a blue ellipse) and B (a rectangle in Fig.~12a, indicating uncertainties in $L_B$ and $T^B_{\rm eff}$) of the components. $T_{\rm eff}^B$, $\sigma_{T_{\rm eff}^B}$ were found from the intersection of the lines $L_B\pm \sigma_{L_B}$ and isochrones chosen based on the position of \mbox{component~A} on the H-R diagram. 
The position of component~A determined by the infrared flux method (IRFM) is shown.

The figure shows that the component masses are: $M_A = 1.66\,M_{\odot}\pm0.04,\ M_B = 0.8\,M_{\odot}\pm0.1$, and the system age is $\log t=9.08\pm0.06$. The main component of $\mu$\,Cet does not belong to the class of giants, but ends its evolution on the main sequence. Note that the sum of the masses $M_A + M_B$ lies within the range from $2.3\,M_{\odot}$ to $2.6\,M_{\odot}$, which is consistent with direct observations.

\section{CONCLUSION}

We have constructed a new orbit of the speckle interferometric pair $\mu$\,Cet based on literature data and 19 observations from this paper, covering 16 epochs from 2018 to 2023. The time of periastron passage during this period determines the high accuracy of the obtained solution. The sum of the system masses amounted to $M_{\rm gaia} = 2.42 \pm 0.05\,M_{\odot}$. 

An analysis of the photometric data from the TESS mission shows that the main component of the system pulsates with the base frequency of $2.0611761(4)$~days$^{-1}$. The values of the base frequency and amplitude are atypical for the $\delta$\,Scuti variables and indicate that this star belongs to the $\gamma$\,Dor  pulsators. This makes $\mu$\,Cet the brightest representative of this class of objects among the stars of the Northern Hemisphere. In addition, we have detected 23 additional frequencies with $S/N>4$. The obtained dependence ``$\Delta P$\,--\,$P$'' demonstrates a downward trend, indicating the prograde pulsation modes.

Two model atmospheres of the A-component of $\mu$\,Cet were constructed using the available spectra. 
The effective temperature and radius of the main component of the system were determined using the photometric data via the infrared flux method. The photometric and spectroscopic data agree with each other. The chemical composition of $\mu$\,Cet was calculated using the two models of its atmosphere. 
No systematic variations in the element abundances with increasing atomic number, 
typical of CP stars, or differences in the abundances relative to the solar abundance of elements within the error were detected, indicating the absence of anomalies in the chemical composition. The abundances of all studied elements in the atmosphere of $\mu$\,Cet are within $\pm0.2$~dex of the solar chemical composition. 
We exclude the possibility that $\mu$\,Cet belongs to the class of magnetic peculiar stars. Unfortunately, the quality of the spectroscopic material is insufficient to detect weak systematic abundance trends that are typical of cool $\delta$\,Scuti stars (for example, in the case of the star 44\,Tau, the excess of heavy elements is 0.3~dex).

\begin{acknowledgments}

The authors thank the National Committee on the Subject of Russian Telescopes \mbox{NKTRT} (\url{https://www.sao.ru/hq/Komitet/}) for allocating the observational time at the BTA telescope. The observations on the SAO RAS telescopes are carried out with the support of the Ministry of Science and Higher Education of the Russian Federation. 
The instrumental base is innovated within the framework of the national project Science and Universities. 
In our study, we used the data from the SIMBAD and VizieR astronomical databases.
This paper includes the data, collected by the TESS mission. Funding for the TESS mission is provided by the NASA's Science Mission Directorate.
The authors thank V.~A.~Vasyuk for implementing the {\tt ORBIT} package in the {\tt SQL} environment.
\end{acknowledgments}

\section*{FUNDING}
The work was carried out within the framework of a grant from the Ministry of Science and Higher Education of the Russian Federation
No.~075--15--2022--262 (13.MNPMU.21.0003).

\section*{CONFLICT OF INTEREST}
The authors declare no conflict of interest.

\begin{flushright}
{\it Translated by M.~Ziazeva}
\end{flushright}

\end{document}